\documentclass[12pt]{article} 
\usepackage{epsfig}
\usepackage{amsfonts}
\usepackage{latexsym}
\usepackage{amsmath,amssymb}
\usepackage{verbatim}
\usepackage{mathtools}
\usepackage{setspace}
\usepackage[all]{xypic}

\usepackage{tikz}
\usepackage[dvipdfm,hypertex]{hyperref}
\usepackage[textheight=9in, textwidth=6.5in, letterpaper]{geometry}
\def\half{{1\over 2}}
\numberwithin{equation}{section}

 \def\p{\partial}

\newcommand{\bea}{\begin{eqnarray}}
\newcommand{\eea}{\end{eqnarray}}
\newcommand{\be}{\begin{equation}}
\newcommand{\ee}{\end{equation}}
\newcommand{\ba}{\begin{align}}
\newcommand{\ea}{\end{align}}

\newcommand{\W}{\mathcal{W}}
\newcommand{\tr}{\mbox{tr}}

\newcommand{\hs}[1]{\mbox{hs$[#1]$}}
\newcommand{\w}[1]{\mbox{$\W_\infty[#1]$}}

\renewcommand{\u}{\mathfrak{u}}

  \makeatletter
  \let\over=\@@over \let\overwithdelims=\@@overwithdelims
  \let\atop=\@@atop \let\atopwithdelims=\@@atopwithdelims
  \let\above=\@@above \let\abovewithdelims=\@@abovewithdelims

\newdimen\tableauside\tableauside=1.0ex
\newdimen\tableaurule\tableaurule=0.4pt
\newdimen\tableaustep
\def\phantomhrule#1{\hbox{\vbox to0pt{\hrule height\tableaurule width#1\vss}}}
\def\phantomvrule#1{\vbox{\hbox to0pt{\vrule width\tableaurule height#1\hss}}}
\def\sqr{\vbox{%
  \phantomhrule\tableaustep
  \hbox{\phantomvrule\tableaustep\kern\tableaustep\phantomvrule\tableaustep}%
  \hbox{\vbox{\phantomhrule\tableauside}\kern-\tableaurule}}}
\def\squares#1{\hbox{\count0=#1\noindent\loop\sqr
  \advance\count0 by-1 \ifnum\count0>0\repeat}}
\def\tableau#1{\vcenter{\offinterlineskip
  \tableaustep=\tableauside\advance\tableaustep by-\tableaurule
  \kern\normallineskip\hbox
    {\kern\normallineskip\vbox
      {\gettableau#1 0 }%
     \kern\normallineskip\kern\tableaurule}%
  \kern\normallineskip\kern\tableaurule}}
\def\gettableau#1 {\ifnum#1=0\let\next=\null\else
  \squares{#1}\let\next=\gettableau\fi\next}

\newcommand{\Tr}{\mbox{Tr}}
\renewcommand{\H}{\mathcal{H}}

\newcommand{\chiu}{\chi^{{\rm U}(\infty)}}
\newcommand{\ff}{\rm f}
\def\half{{1 \over 2}}

\def\tr{{\rm Tr}}

\def\Or[#1]{{\text{O}}\left({#1}\right)}
\def\dotl[#1,#2]{\left\langle #1, #2 \right\rangle}
\def\dotlb[#1,#2]{[ #1, #2 ]}
\def\dotp[#1,#2]{(#1) \cdot (#2)}
\def\aff[#1,#2]{\hat{#1}(#2)}
\def\n4sym{{\cal N}=4 SYM}
\def\>{\rangle}
\def\<{\langle}
\def\weight[#1,#2,#3]{\{(#1),#2,#3\}}
\def\ads[#1]{$\text{AdS}_{#1}$}

\begin{document}

\ \\

\begin{center}

{ \LARGE {\bf  Partition Functions of \\ 
\vspace{.5cm}
Holographic Minimal Models}}

\vspace{0.8cm}

Matthias R. Gaberdiel$^{a}$, Rajesh Gopakumar$^b$, Thomas Hartman$^c$, and Suvrat Raju$^b$
\vspace{1cm}

{\it $^a$Institut f\"ur Theoretische Physik, ETH Zurich, \\
CH-8093 Z\"urich, Switzerland \\
\tt{\small gaberdiel@itp.phys.ethz.ch}
 }

\vspace{0.5cm}

{\it $^b$Harish-Chandra Research Institute, \\
Chhatnag Road, Jhunsi, Allahabad, India 211019\\ 
\tt{\small gopakumr@hri.res.in, suvrat@hri.res.in}
}

\vspace{0.5cm}

{\it $^c$ Institute for Advanced Study, School of Natural Sciences,
\\
Princeton, NJ 08540, USA \\ 
\tt{\small hartman@ias.edu}
}

\vspace{1.0cm}

\end{center}

\begin{abstract}
The partition function of the $\W_N$ minimal model CFT is computed in the large $N$ 't~Hooft limit
and compared to the spectrum of the proposed holographic dual, a 3d higher spin gravity 
theory coupled to massive scalar fields. At finite $N$, the CFT contains additional light states that
are not visible in the perturbative gravity theory. We carefully define the large $N$ limit, and 
give evidence that, at $N =\infty$, the additional states become null and decouple from all correlation 
functions. The surviving states
are shown to match precisely (for all values of the 't~Hooft coupling) 
with the spectrum of the higher spin gravity theory. 
The agreement between bulk and boundary is partially explained by 
symmetry considerations involving the conjectured equivalence between 
 the $\W_N$ algebra in the large $N$ limit and the higher spin algebra 
of the Vasiliev theory.
\end{abstract}

\pagestyle{empty}

\pagebreak
\setcounter{page}{1}
\pagestyle{plain}

\setcounter{tocdepth}{2}

\tableofcontents

\section{Introduction}
It is rare to find an example of a holographic duality where we can compute physical 
quantities on both sides at the same point in parameter space. Early checks of the AdS/CFT 
correspondence relied on matching quantities  that were protected, such as supersymmetric 
partition functions, and so could be computed at weak coupling both in gravity and in the CFT.  
The past few years have seen impressive progress beyond this using techniques such as 
integrability \cite{Beisert:2010jr} and supersymmetric localization  \cite{Pestun:2007rz}
that have allowed us to compute the coupling-constant dependence of some quantities 
all the way up to strong coupling. However, these techniques crucially require large $N$ or 
supersymmetry, and as a consequence some of the most 
interesting questions about quantum gravity, like the information-loss puzzle, have so far remained 
out of  reach in string realizations of AdS/CFT.

However, if we {\em can} find the quantum gravity dual of an exactly solvable field theory, then the 
AdS/CFT correspondence opens the possibility to understand quantum gravity 
quantitatively since the puzzles of quantum gravity can then perhaps be mapped to calculable 
results in field theory.
The simplest field theory is, of course, that of free fields.  Klebanov and Polyakov \cite{Klebanov:2002ja} (see also 
\cite{Witten,Mikhailov:2002bp,Sezgin:2002rt} for earlier work) conjectured that the free 
${\rm O}(N)$ vector model in $d=3$ is dual to Vasiliev's theory of higher spin gravity in AdS$_4$ 
\cite{Vasiliev:2003ev} (see for example
\cite{Vasiliev:1999ba,Bekaert:2005vh,Iazeolla:2008bp,Campoleoni:2009je} for reviews), and that the interacting vector model at its critical point is dual to the same bulk theory with alternate boundary conditions; for recent progress on this proposal, see 
\cite{Giombi:2009wh,Giombi:2010vg,Koch:2010cy,Giombi:2011ya}.  

In $d=2$, there is a rich landscape of solvable interacting CFTs.  One family of such CFTs is that of 
the $\W_N$ minimal models, a generalization of the $c<1$ Virasoro minimal models that admits a large $N$ limit. Following the demonstration \cite{Henneaux:2010xg,Campoleoni:2010zq,Gaberdiel:2010ar} that 3d higher spin gravity has $\W$ symmetry, it was conjectured in \cite{Gaberdiel:2010pz} that the large $N$ limit of the $\W_N$ minimal model is dual to a particular AdS$_3$ higher spin theory of Vasiliev \cite{Prokushkin:1998bq,Prokushkin:1998vn}. As evidence, it was shown that the first few low-dimension operators in the CFT match the lightest states in the bulk.  Furthermore, an RG flow relating different fixed points of the CFT was found to match the behaviour in the bulk.  In 
\cite{Gaberdiel:2011wb}, the asymptotic symmetries of the bulk theory were computed and shown to agree with the CFT, at least in certain special cases where both sides can be computed explicitly.
Very recently, the 
generalization to the SO case has been discussed in \cite{Ahn:2011pv,GV}. 

In this paper we demonstrate an important piece of this conjecture in the large $N$ limit.  Specifically, we compute the leading large $N$ CFT partition function, and find it to be equal to the bulk 1-loop determinant, under several assumptions spelled out below.  Therefore the full CFT partition function contains the entire perturbative spectrum of the higher spin gravity theory.  This is a statement about the 
large $N$ limit, and it relies on subtleties of how the limit is defined; finite $N$ remains an open question and will require new ingredients as described below.  The CFT, of course, can be defined and solved for any $N$, so ideally the duality will serve as a guide towards formulating the quantum Vasiliev theory.

We begin by describing the general features of the bulk and boundary spectrum. The bulk theory 
has an infinite tower of massless fields with spins $s=2,3,\dots$ coupled to two complex scalars
$\phi_\pm$. Interactions among the higher spin fields are dictated by the higher spin Lie algebra, 
which has a free parameter $\lambda$ in its commutation relations.  The scalars have equal mass 
fixed by the algebra \cite{Vasiliev:1992ix,Prokushkin:1998bq,Prokushkin:1998vn},\footnote{The mass 
condition on the scalar is sufficient to specify the theory to cubic order. At higher orders, the theory 
with two scalars (unlike that of one scalar) is only determined up to various discrete choices. We 
thank X.\ Yin for comments on this point.}

\be
M^2 = -1 + \lambda^2 \ ,
\ee
but they are quantized with opposite (conformally invariant) boundary conditions and therefore the corresponding conformal weights are
\be\label{hpm}
h_+ = \frac{1+\lambda}{2} \ , \qquad h_- = \frac{1-\lambda}{2} \ .
\ee
The full perturbative spectrum in the bulk is simple:  it consists of the scalars $\phi_\pm$, boundary excitations of the higher spin fields (generalizing the Brown-Henneaux boundary gravitons), and multiparticle states with any number of $\phi_+$, $\phi_-$, and higher spin excitations.

The dual $\W_N$ CFT is the coset theory
\begin{equation}\label{gencos}
\frac{\mathfrak{su}(N)_k \oplus \mathfrak{su}(N)_1 }{ \mathfrak{su}(N)_{k+1}} \ ,
\end{equation}
so it is labeled by two positive integers $N$ and $k$ (see \cite{Bouwknegt:1992wg} for a review). 
The bulk mass parameter $\lambda$ is identified with the boundary 't~Hooft coupling, defined by
\be
\lambda = \frac{N}{N+k} \ ,
\ee
which is held fixed in the large $N$ limit. The spectrum at finite $N,k$ is known exactly: up to 
field identifications primary states are 
labeled by two Young tableaux denoting integrable weights $(\Lambda_+; \Lambda_-)$ of 
$\mathfrak{su}(N)_k$ and $\mathfrak{su}(N)_{k+1}$, respectively. Their conformal dimension equals
\be
h_{(\Lambda_+;\,\Lambda_-)} = 
 \frac{1}{2p(p+1)}\left( \big| (p+1)\Lambda_+ - p\Lambda_- + \rho\big|^2  - \rho^2\right)\ ,
\ee
where $p \equiv k+N$ and $\rho$ is the Weyl vector of $\mathfrak{su}(N)$.  At large but finite $N$, the CFT 
spectrum  has many closely spaced light states with $h \ll N$.  These are states where the two Young 
tableaux differ by only ${\cal O}(1)$ boxes. For example, as $N\rightarrow \infty$, the CFT naively becomes
infinitely  degenerate because every state with $\Lambda_+ = \Lambda_-$ has conformal dimension
$h_{(\Lambda;\Lambda)}=\tfrac{C_2(\Lambda)}{p(p+1)}$, which goes to zero provided that 
the quadratic Casimir $C_2(\Lambda)$  grows slower  than $N^2$.
However, the representation theory changes qualitatively as $N\rightarrow \infty$, as a `primary' state at 
finite $N$ will not necessarily define a primary state of an irreducible representation in the limit.  We will 
show that, as a result, many of these light states decouple from the theory as $N\rightarrow \infty$, so we 
must be careful about how the limit is defined and what states are included. 

In order to explain the relevant phenomena, it is useful to identity different classes of CFT states as 
$N\rightarrow \infty$.  First we have those states where both $\Lambda_+$ and $\Lambda_-$ can 
be obtained in finite tensor powers of the fundamental ($\ff$) or antifundamental ($\bar\ff$) representation;
the corresponding Young tableaux then have finitely many boxes and `antiboxes'. The simplest such states are the
primaries where either $\Lambda_\pm$ is just the fundamental $\ff$ or antifundamental $\bar\ff$ representation. In the large
$N$ limit, the conformal dimensions of these generating fields are 
\be
h_{(\ff; 0)} = h_{(\bar\ff;0)} = h_+ \ , \qquad h_{(0;\ff)} = h_{(0;\bar\ff)} = h_-  \ ,
\ee
with $h_{\pm}$ as defined in eq.~(\ref{hpm}).
All other primaries with finitely many boxes and antiboxes can be obtained by taking finite
tensor powers of these primitive states. Many of them  can be directly matched to certain 
multiparticle states in the bulk, without any extra complications. However, there is a subtlety 
whenever both $\Lambda_+$ and $\Lambda_-$ contain non-trivial powers of the fundamental 
representation (or both contain non-trivial powers of the anti-fundamental representation). 
The simplest example appears for $(\ff; \ff) = (\ff; 0) \otimes (0; \ff)$, for which the dimension of the
corresponding `multiparticle' state seems to vanish as $N\rightarrow \infty$. Moreover, it does not have 
the additive conformal dimension that is the hallmark of large $N$ gauge theories with gravity duals, 
\be
h_{(\ff; \ff)} = {\cal O}(\tfrac{1}{N}) \neq h_+ + h_- = 1\ .
\ee
Since we are comparing the CFT spectrum to multiparticle states of a free field in the bulk for 
which energies are obviously additive, this seems like a contradiction,  but it is not.  The reason
is that the CFT  representation labeled by $(\ff; \ff)$ contains a \textit{descendant} state with the appropriate 
conformal dimension $h = h_+ + h_-$.  
We will show that in our large $N$ limit (for $\lambda \neq 0$) the descendant becomes the 
generating state of the representation, while the original primary state becomes null.
Therefore the state at $h \sim 1/N$ 
decouples from all correlation functions and is
dropped from the spectrum, leaving only the 
representation generated from the state 
at $h=1$ that agrees with the bulk.  A similar situation arises for any representation where both 
$\Lambda_+$ and $\Lambda_-$ contain fundamentals, or both contain antifundamentals, such as 
$(\ff ; \tableau{2})$, $(\bar{\ff}\ff ; \tableau{1 1})$, {\it etc}.  We study several examples of such representations 
and find that the states which survive as $N\rightarrow \infty$ are always those that agree with 
the gravity prediction. However we do not have a general proof that this holds for all states in this class;
our result therefore relies on the conjecture that this structure continues for higher representations.  

The representations just discussed -- light states 
formed by ${\cal O}(N^0)$ tensor powers of fundamentals and antifundamentals -- precisely account for the 
perturbative gravity spectrum.  Demonstrating this fact is the main goal of this paper.  

We should note that there is another potentially interesting class of states.
These are the states 
where either $\Lambda_+$ or $\Lambda_-$ (or both) have ${\cal O}(N)$ (or greater) boxes or antiboxes. 
Generically, these states have conformal dimension $h\sim N$ or greater, but there are also
light states with $h\ll N$ among them: they arise if $\Lambda_+$ and $\Lambda_-$ only differ by 
${\cal O}(1)$ boxes and antiboxes. An unusual feature of these states 
(from the point of view of AdS/CFT) is that they have a density of states 
growing exponentially with $N$. There may be a large $N$ limit of the CFT in which these
latter states are included as a continuum, but they are not included in our limit, and they do not 
correspond to perturbative  gravity states.
(See \cite{Runkel:2001ng,Roggenkamp:2003qp} for a similar choice in the 
$c\rightarrow 1$ limit of the Virasoro minimal models.) In any case, any such states are decoupled 
from the perturbative gravity states because the fusion rules of the CFT do not allow two states 
with a fixed finite number of boxes and antiboxes to produce states where this number grows with 
$N$. Note that this decoupling is qualitatively different in nature from the decoupling phenomenon 
described above.

All of our results pertain to the $N\rightarrow \infty$ limit, but since ultimately the aim is to address 
quantum gravity and black holes, let us briefly make a few comments about finite $N$. (We return to 
this point in the discussion section.) At finite $N$, 
the distinction between the above classes of states gets blurred. 
Since the number of boxes are essentially preserved by the fusion rules, there will still be 
a natural separation between the states made from a small number of Young tableaux boxes, and
the exponentially growing number of states with ${\cal O}(N)$ boxes. However the decoupling of light 
null states with $\mathcal{O}(N^0)$ boxes that 
was crucial for matching to the 1-loop gravity answer will no longer occur.  
It would be interesting to determine how strongly these new light states couple to
other perturbative states at finite $N$, and, in particular,  whether their contribution 
to four-point functions is suppressed beyond the usual $1/N$ suppression of connected 
correlation functions. If no additional suppression occurs, the bulk theory will need to be modified
in order to reproduce the leading $1/N$ corrections in the CFT via tree-level Witten diagrams.
\medskip

The paper is organized as follows.
The matching of the perturbative bulk spectrum with that of the large $N$ CFT 
is outlined in section \ref{secsummary}, and given 
in detail in sections \ref{s:gravity} - \ref{s:fusion}.  The basic logic is as follows.  In section \ref{s:gravity}, 
we rewrite the bulk 1-loop partition function as a sum over characters of ${\rm U}(\infty)$ to facilitate 
comparison with the CFT.  
In section \ref{s:cft}, we take the large $N$ limit of the exact CFT partition function.  The result is naturally 
written in terms of the modular $S$-matrix of ${\rm SU}(N)$ Wess-Zumino-Witten theory,
and the quantum dimensions of $\Lambda_+, \Lambda_-$, which can be rewritten as ${\rm U}(\infty)$ 
characters similar to the bulk answer.\footnote{Interestingly, many results in this section can be easily 
obtained from the topological string literature on knot invariants by replacing 
$q_{\rm there} = e^{2\pi i\over k+N} \rightarrow q_{\rm here} = e^{2 \pi i \tau}$, with 
$\tau$ the complex structure of the torus.} Next, the structure of the degenerate representations 
is described in more detail, decoupled null states are subtracted from the partition function, and the 
CFT answer is then 
shown to  match exactly the gravity answer.  In section \ref{s:fusion}, we justify this last 
step of subtracting null states by deriving the structure of the CFT representations from a fusion analysis.
This leads to the picture described above, where some primaries become null and descendants take their 
place.

Section \ref{s:hsalgebra} addresses the role of the higher spin algebra in the spectrum of the two theories, 
and stands independently from the rest of the paper.  
It is shown that the ${\rm U}(\infty)$ characters that 
appear in the partition functions agree with the characters of the bulk higher spin 
algebra \hs{\lambda}, and thus we can reinterpret both partition functions as a sum over characters of 
\hs{\lambda}. As in \cite{Gaberdiel:2011wb}, it is argued from an algebraic point of view that the 't Hooft 
large $N$ limit of the $\W_N$ algebra contains the bulk higher spin algebra \hs{\lambda}, and further evidence 
for this claim is given.  This conjectured equivalence of symmetries thus explains why also the boundary 
partition function should have an \hs{\lambda} symmetry. The duality, however, goes beyond the
requirements of symmetry, because not only the form of the characters 
but the choice of representation content in the bulk and boundary agree.

Finally, in section \ref{s:discussion}, we discuss open questions and conclude.  Some conventions and 
formulae for ${\rm SU}(N)$ characters and Schur polynomials are given in appendix \ref{a:orthobasis}; details of 
the fusion calculations appear in appendix \ref{a:fusiondetails}; and additional evidence for the role of the 
higher spin algebra is described in appendix \ref{a:hsdetails}.

\section{Summary of Results \label{secsummary}}

The conclusions of this work rest on a number of technical computations which we will present in detail in the following sections (and appendices). Here we will give the reader a summary of the overall logic of the work. We spell out our working assumptions together with the precise statements of results we can demonstrate as well as those which we conjecture based on strong evidence. For the proofs (of assertions) and worked out examples (in support of conjectures) we then refer the reader to the appropriate sections. 

\subsection{The Bulk Spectrum}
 
As described in the introduction, we will match the spectrum of quadratic fluctuations 
in the higher spin theory on AdS$_3$ (with the two complex scalars) with the 
corresponding answer from the CFT side to leading order in the large $N$ 't Hooft limit 
(and for $0<\lambda <1$). The expression on the gravity side 
is relatively simple and hence we begin our analysis with this as our starting point.

The spectrum of perturbative physical excitations in the higher spin theory is obtained from the computation of the one loop determinant of the kinetic terms of all the fields in the theory (appropriately gauge fixed). In our case, we have a massless sector comprised of  gauge fields with spins $2,3,\dots$ as well as 
two propagating complex scalar fields of equal mass,
\be
\label{mscalar}
M^2 = -1 + \lambda^2  \ ,
\ee
where $0 < \lambda < 1$.
The two scalars are, however, quantized with opposite (conformally invariant) boundary 
conditions, and  the corresponding conformal weights are as given  in eq.~(\ref{hpm}). 
The (generalized) Brown-Henneaux central charge of the theory is equal to 
$c={3\ell \over 2G_N}=(N-1)(1-\lambda^2)$. The last parametrization is chosen so as to 
agree with the CFT central charge.\footnote{One of the reasons for not considering 
$\lambda=1$ is the vanishing of the leading piece of the central charge. Another is the 
divergence from one of the scalar factors in eq.~(\ref{bulkz}) since $h_-=0$.}

The one loop partition function on thermal AdS$_3$ was evaluated in
\cite{Gaberdiel:2010ar} using heat kernel techniques 
\cite{Giombi:2008vd,David:2009xg}. Combining this with the classical action 
$Z_{\rm cl}=(q\bar{q})^{-c/24}$ on the solid torus with boundary modular parameter $q = e^{2\pi i\tau}$ we obtain
\be\label{bulkz}
Z_{\rm bulk} = Z_{\rm cl}\cdot Z_{\rm 1-loop}= (q\bar{q})^{-c/24}\cdot Z_{\rm hs} \cdot
Z_{\rm scal}(h_+)^2 \cdot Z_{\rm scal}(h_-)^2 \ .
\ee
Here
\be\label{macdef}
Z_{\rm hs} = \prod_{s=2}^{\infty}\prod_{n=s}^\infty \frac{1}{|1-q^n|^2} 
=\prod_{n=2}^{\infty}\frac{1}{|(1-q^n)^{n-1}|^2} 
\equiv |\tilde{M}(q)|^2 
\ee
is the contribution of the generalised boundary gravitons of the higher spin fields. This matches 
with the vacuum $\W_N$ character \cite{Gaberdiel:2010ar}. Note that $\tilde{M}(q)$ essentially equals
the MacMahon function. 

Each real scalar contributes 
\be\label{scalar1intro}
Z_{\rm scal}(h) = \prod_{j,j'=0}^\infty\frac{1}{1-q^{h+j}\bar{q}^{h+j'}} \ ,
\ee
and hence the contribution of a complex scalar is the square of (\ref{scalar1intro}). 
Its form can be understood intuitively: a local operator of dimension $h$ has 
descendants which are obtained by acting on it with derivatives. Thus the `single particle' contribution to 
the  partition function is given by
\begin{equation}
Z_{\rm sing\, par}(h,q,\bar{q}) = {q^h \bar{q}^h \over (1 - q)(1 - \bar{q})}\ . 
\end{equation}
In the non-interacting limit, where we can neglect the anomalous dimensions of
composite operators, we can obtain the `multi-particle' partition function 
by using the standard formula for Bose statistics, leading to 
\be\label{scalar1}
Z_{\rm scal}(h) = \exp{\left[\sum_{n=1}^{\infty} {Z_{\rm sing\, par}(h,q^n, \bar{q}^n) \over n}\right]} 
=  \prod_{j,j'=0}^\infty\frac{1}{1-q^{h+j}\bar{q}^{h+j'}} \ .
\ee
\medskip

To compare with the CFT and to exhibit more transparently the nature of the multiparticle states in the 
bulk we first rewrite the scalar determinant in terms of U$(\infty)$ characters.  Characters of $\u(N)$ in 
a representation $R$ are given by Schur polynomials in $N$ variables,
\be
\chi_R^{\u(N)}(z_i) = P_R(z_i) \ , \qquad i=1,\ldots, N \ .
\ee
Taking the large $N$ limit and evaluating on the Weyl vector, we define the specialized Schur functions
\bea\label{schurpm}
P_R(q) &\equiv& \chi_R^{\mathfrak{u}(\infty)}(z_i) \ ,  \quad (z_i = q^{i - \half}) \ , \\
P_R^\pm(q) &\equiv& q^{\pm \frac{\lambda}{2} B(R)} P_R(q) \ , \notag
\eea
where $B(R)$ is the number of boxes in the Young tableau $R$. Explicit formulae for the Schur functions 
are given in appendix \ref{a:orthobasis}. 
In terms of U$(\infty)$ characters, the scalar determinant (\ref{scalar1intro}) is
\be\label{scalpoly}
Z_{\rm scal}(h_\pm) = \sum_{R} |P_R^{\pm}(q)|^2 \ .
\ee
Here the sum is over all Young tableaux of ${\rm U}(\infty)$, {\it  i.e.}\ without any restrictions on the 
lengths of rows or columns; the proof of this statement is given in Sec.~3. Combining the contribution
of the two complex scalars we therefore obtain
\be\label{gravityfinal1}
Z_{\rm bulk} =  (q\bar{q})^{-c/24}\cdot |\tilde{M}(q)|^2\cdot 
\sum_{R_+,S_+,R_-,S_-}| P_{R_+}^+(q) P_{S_+}^+(q) P_{R_-}^-(q) P_{S_-}^-(q) |^2 \ .
\ee
In this form it is natural to view the gravity answer as the combined contribution from (weakly coupled) multi-particle states of the complex scalar with dimension $h_+$ (the pieces $R_+$, $S_+$), and that of the scalar with dimension $h_-$ (the pieces $R_-$, $S_-$) all dressed with the boundary graviton excitations in
$\tilde{M}(q)$.

\subsection{The CFT spectrum}\label{s:sumCFT}

We wish to compare $Z_{\rm bulk}$ in eq.~(\ref{gravityfinal1}) with the diagonal modular invariant 
partition function of the $\W_N$ minimal model at leading order in the large $N$ 't Hooft limit. 
This turns out to be somewhat subtle for the following reason. Irreducible Virasoro representations 
of the $\W_N$ theory for finite $N$, $k$ often become reducible in the large $N$ limit, and characters go 
over to sums of characters of the reducible blocks. When this happens, some components of the 
reducible representation can decouple from all correlation functions, much like ordinary null states.  
We therefore need to make a correction to the naive partition function (which includes all these states) 
that removes the contribution from these decoupled states. We argue that when this is correctly accounted 
for,  the CFT answer matches exactly with the gravity answer in eq.~(\ref{gravityfinal1}).

The CFT partition function (for any finite $N$, $k$) is a  sum over characters (\textit{i.e.}, branching functions) 
of the defining coset theory,
\be\label{zcftschem}
Z_{\rm CFT}(N,k) = \sum_{\Lambda_+, \Lambda_-} |b_{(\Lambda_+;\Lambda_-)}(q)|^2 \ ,
\ee
where $(\Lambda_+,\Lambda_-)$ are allowed representations of $\mathfrak{su}(N)_k$ 
and $\mathfrak{su}(N)_{k+1}$, respectively. The branching functions (at finite $N$, $k$) are known
explicitly (see eq.~(\ref{brfn})). 

\begin{figure}
\begin{center}
\epsfig{file=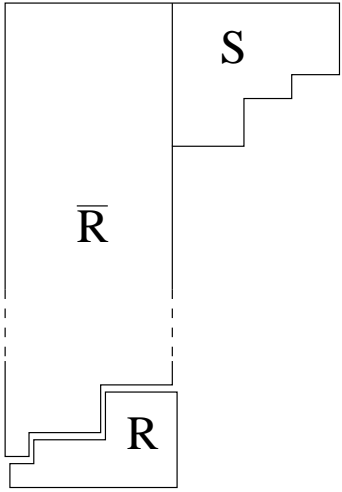}
\caption{\small A Young tableau of SU$(N)$ in the large $N$ limit.  The full representation $\Lambda = (\overline{R}, S)$ has a finite number of `boxes'  $S$ and `antiboxes' $R$.\label{fig:tableaux}}
\end{center}
\end{figure}

For the comparison with gravity, as we take the $N\rightarrow \infty$ limit, we shall only 
consider those representations $\Lambda_\pm$ that are contained in finite tensor powers
of the fundamental and anti-fundamental (where the number of tensor powers does not scale with $N$).
The corresponding Young tableaux can then be viewed as  two Young tableaux 
placed side by side,
\be
\Lambda_\pm = (\overline{R}_\pm, S_\pm) \ ,
\ee
where $\overline{R}_\pm$ is a tensor power of anti-fundamentals (`antiboxes') and 
$S_\pm$ is a tensor power of fundamentals (`boxes') as in fig.~\ref{fig:tableaux}.  

In the large $N$, $k$ 't~Hooft limit, the branching functions for these representations simplify considerably, 
and can be written as 
\bea
\label{brfnsimp1overv}
b_{(\Lambda_+;\Lambda_-)}(q) & \cong &
q^{-\frac{N-1}{24}(1-\lambda^2)}\, \tilde{M}(q) 
q^{\frac{\lambda}{2} (B_+-B_-)} \, q^{C_2(\Lambda_+) + C_2(\Lambda_-)}\, 
\frac{\sum_{w \in W} \epsilon(w)\, q^{-\langle w(\Lambda_+ + \rho), \, (\Lambda_- + \rho)\rangle }}
{\sum_{w \in W} \epsilon(w) \, q^{-\langle w(\rho), \rho\rangle} }\, \cr
& \cong &
q^{-\frac{c}{24}}\, \tilde{M}(q) \, q^{\frac{\lambda}{2} (B_+-B_-)} \, q^{C_2(\Lambda_+) + C_2(\Lambda_-)}\,
\frac{S_{\Lambda_+\Lambda_-}}{S_{00}}  \ .
\eea
Here $C_2(\Lambda)$ is the quadratic Casimir while 
\be\label{Bpm}
B_{\pm}=B(\Lambda_{\pm}) \equiv B(R_\pm) + B(S_\pm)
\ee 
denotes the sum of the number of boxes and antiboxes.  The sums in the first line of (\ref{brfnsimp1overv}) 
are over the Weyl group of the finite dimensional Lie algebra $\mathfrak{su}(N)$; in the second line 
they are written in terms of the modular $S$-matrix of the affine $\mathfrak{su}(N)$ algebra.  Note that
the modified MacMahon function $\tilde{M}(q)$ (corresponding to the higher spin contribution to the bulk
answer) appears as a common factor in each of the branching functions.  

Using the Verlinde formula we can relate the branching function for $(\Lambda_+;\Lambda_-)$ 
in the large $N$, $k$ limit to  branching functions  associated to representations of the form $(\Lambda;0)$
\be\label{generalbranch1}
b_{(\Lambda_+;\Lambda_-)}(q)  \cong q^{\frac{\lambda}{2}(B_+ - B_-)}
\sum_{\Lambda} N_{\Lambda_+\overline{\Lambda}_-}{}^{\Lambda} \,  
q^{-\frac{\lambda}{2} B(\Lambda)} \, b_{(\Lambda;0)}(q) \ .
\ee
In this limit, the fusion coefficients $N_{\Lambda_+\overline{\Lambda}_-}{}^{\Lambda}$ 
are just ${\rm SU}(\infty)$  tensor product multiplicities for the representation 
$\Lambda$  in  $\Lambda_+\otimes \overline{\Lambda}_-$. 
Further, we can rewrite $b_{(\Lambda;0)}(q)$ in terms of the quantum dimension
\be
b_{(\Lambda;0)}(q) \cong  \, q^{-\frac{N-1}{24}(1-\lambda^2)}\cdot \tilde{M}(q) \cdot
q^{\frac{\lambda}{2} B(\Lambda)}\,     q^{C_2(\Lambda)} \cdot  \dim_q(\Lambda)   \ . 
\ee
This immediately provides a link to the characters of ${\rm U}(\infty)$ that appeared in the scalar 
contribution to the bulk answer (\ref{scalpoly}). In particular, we have 
\bea\label{uinfchar}
b_{(\Lambda;0)}(q) &\cong& q^{-\frac{N-1}{24}(1-\lambda^2)}\cdot    \tilde{M}(q)  \cdot
q^{\frac{\lambda}{2} B(\Lambda)}\cdot P_{R^T}(q) \, P_{S^T}(q)     \\
&\cong& q^{-\frac{N-1}{24}(1-\lambda^2)}\cdot   \tilde{M}(q)  \cdot
P_{R^T}^+(q) \, P_{S^T}^+(q)   \  ,
\eea
where we have labelled $\Lambda= (\overline{R}, S)$  as in fig. \ref{fig:tableaux}, and $R^T$ and $S^T$ 
indicate the transposed representations where the Young tableaux are flipped along the diagonal.   

The form of the general branching function eq.~(\ref{generalbranch1}) suggests that some of the
representations become reducible in the large $N$, $k$ limit. For example for 
$\Lambda_+=\Lambda_-={\rm f}$, the fundamental, the branching function of $({\rm f};{\rm f})$
equals the sum of the branching functions associated to $(\Lambda;0)$, where
$\Lambda$ is either the trivial or the adjoint representation --- the two representations
that appear in the tensor product $({\rm f}\otimes\bar{\rm f})$
\be\label{simplesplit1over}
b_{(\ff;\ff)} = q^{-\frac{c}{24}}(1 + q^2 + \cdots) + q^{-\frac{c}{24}}(q + 2 q^2 + \cdots) \ . 
\ee
However, as we will see in detail in
sec.~5, the resulting representation cannot be decomposed into a direct sum of two irreducible 
representations. Instead it is indecomposable, and its structure is described by 
\be\label{ffdia}
 (\ff;\ff): \qquad 
\xymatrix@C=1pc@R=2.2pc{
   & \vdots & & \vdots & & \vdots & & \vdots \\
   & 2 & & \ar@<0.4ex>[dr]^{L_1} \rho & & \ar@<-0.4ex>[dl] \xi & & \ar@<-0.4ex>[lldd]_{L_2} T \\
  & 1 & & & \psi\ar@<0.4ex>[ul]^{L_{-1}} \ar@<-0.4ex> [ur] \ar[dr]_{L_1}& & &\\
 L_0 =& 0 & &  & & \omega\ar@<-0.4ex>[uurr]_{L_{-2}} & & 
}
\ee 
$\omega$ and $\psi$ correspond to the leading terms of the first and second bracket 
in (\ref{simplesplit1over}), respectively; 
$\psi$ and its descendants, such as $\rho$ and $\xi$, matches the gravity contribution, while 
the representation built on $\omega$ is the extra piece.  From the arrows in the diagram 
it is clear that the representation is indecomposable (all states are connected) and that $\psi$ is the cyclic 
state (any state can be reached starting from $\psi$). One can show that $\omega$ and its descendants 
decouple from correlation functions. We can 
therefore  drop these states from the limiting spectrum 
in the large $N$ limit. The states that survive (namely, $\psi$ and its descendants) exactly match the 
gravity prediction.  

This pattern appears to continue. In the tensor product $\Lambda_+ \otimes \overline{\Lambda}_-$ the 
only $\Lambda$ which contribute ({\it i.e.}\ which do not decouple) are the ones where no boxes and 
antiboxes are annihilated into singlets. (In the above example, the vacuum representation is an example 
where a box and an antibox annihilate.) We give evidence for this pattern in the explicit calculations of 
sec.~4 where we have worked out some non-trivial examples for which we can explicitly check this. 
We will therefore assume this to be true though it would be very nice to have a general proof of this fact. 
As mentioned before, this decoupling appears to happen only for $\lambda\neq 0$. It would be interesting 
to understand better the nature of the $\lambda=0$ theory.

The corrected contribution of the branching functions to the CFT partition function is thus  given by 
considering only those $\Lambda$ in  eq.~(\ref{generalbranch1}) for which 
\be
B(\Lambda) = B(\Lambda_+) + B(\Lambda_-) \ .
\ee
With this restriction, the corrected branching function equals
\be\label{fcft}
{\rm ch}^{\rm cft}_{R_+ S_+ R_- S_-}(q) = q^{-{c\over 24}}\cdot \tilde{M}(q)
\cdot P_{R_+^T}^+(q) P_{S_+^T}^+(q) P_{R_-^T}^-(q) P_{S_-^T}^-(q) \ .
\ee
Thus the CFT answer in the strict infinite $N$ limit is given by 
\be\label{cftanswer}
Z_{\rm CFT}(\lambda) = \sum _{R_+S_+R_-S_-} |{\rm ch}^{\rm cft}_{R_+ S_+ R_- S_-}(q)|^2 \ .
\ee
We see that this is precisely the bulk answer as given in eq.~(\ref{gravityfinal1}).  

The ${\rm U}(\infty)$ characters that appear here were introduced as a compact way of keeping 
track of $q$'s in the partition function, with no apparent physical origin. In fact these characters 
have a natural interpretation: they are characters of the higher spin algebra \hs{\lambda} governing 
the bulk theory.  Indeed, representations of \hs{\lambda} are labeled by Young tableaux, and
\be
P_R^\pm(q) = \Tr_R q^{L_0} \ ,
\ee
where the right-hand side is a trace in \hs{\lambda}. Here $L_0$ is the conformal weight, 
which appears as the zero mode of $\mathfrak{sl}(2)\subset\hs{\lambda}$.  We will argue in sec.~\ref{s:hsalgebra} 
that characters of \hs{\lambda} appear naturally in the 't~Hooft limit of the CFT, because $\W_N$ contains 
a higher spin subalgebra in the large $N$, large $c$ limit.  This subalgebra consists of the 
`wedge' modes  $W^s_m$ with $|m|<s$, where $s$ is the spin; 
this is analogous to the $\mathfrak{sl}(2)$ subalgebra of the Virasoro algebra.  Thus in (\ref{fcft}), the Schur 
polynomials are characters of the higher spin subalgebra, while $\tilde{M}(q)$ accounts for 
descendants obtained by acting with modes outside the wedge, $|m| \geq s$.  Although this 
provides a clear physical interpretation of the quantities appearing in the partition functions, 
it requires more machinery to explain fully, relies on a well supported but unproven conjecture 
about the algebras, and is not required to prove 
$Z_{\rm CFT} = Z_{\rm bulk}$, so we postpone further discussion to section \ref{s:hsalgebra}.

\section{Gravity Partition Function}\label{s:gravity}

The bulk side of the duality proposed in \cite{Gaberdiel:2010pz} is a theory of higher spin gravity coupled to two massive scalars.  The massless sector, with all spins $2,3,\dots$, can be described as a Chern-Simons gauge theory based on two copies of the infinite-rank higher spin Lie algebra \hs{\lambda}, a generalization of su$(\infty)$.  As we have described in section \ref{secsummary}, the higher spin sector is coupled to two propagating complex scalar fields that
have equal masses, given by \eqref{mscalar}, but are quantized with different boundary conditions, making them dual to operators of conformal dimension given by \eqref{hpm}. Combining with the contribution of the higher spin fields \cite{Gaberdiel:2010ar} leads to the one-loop partition function \eqref{bulkz} \cite{Gaberdiel:2010pz}, where we recall that $Z_{\rm scal}(h)$ is given by \eqref{scalar1intro} \cite{Giombi:2008vd,David:2009xg}.

Our goal is to match the full 1-loop partition function \eqref{bulkz} to the partition function of the dual minimal model in the strict large $N$ limit.  First we will rewrite the gravity partition function so that it more closely resembles the CFT.  This can be achieved by rewriting the scalar contribution in terms of ${\rm U}(N)$ characters and using standard manipulations of orthogonal polynomials (see for example \cite{Macdonald,FultonHarris}).  The theory does not have any obvious ${\rm U}(N)$ symmetry, so for now this can be thought of as a bookkeeping device; we will return to the interpretation below and in section \ref{s:hsalgebra}.  Let $U=\mbox{diag}(z_1,\dots,z_N)$ be a diagonal 
${\rm U}(N)$ matrix and $R = (r_1 , \dots, r_N)$ label a Young tableau with rows of length $r_i$.  The trace in the representation $R$ is a Schur polynomial,
\be\label{schurchar}
\Tr_R U = \frac{\det z_i^{r_j + N - j}}{\det z_i^{N-j}}  \ .
\ee
We observe that\footnote{The expression (\ref{scalarzOV}) is similar to the Ooguri-Vafa operator  \cite{Ooguri:1999bv} (see also \cite{Labastida:2000zp,Labastida:2000yw,Marino:2004uf}) that arises in the topological string computation of knot invariants. There it arose from integrating out a scalar field living on a knot in $S^3$, whereas here the scalar lives in a solid torus; the detailed operators are different but the group structure is the same, so the manipulations applied in that context are useful here.}
\be\label{scalarzOV}
Z_{\rm scal}(h) = \exp\left[\sum_{n=1}^\infty \frac{1}{n} \Tr_{\ff} U^n \, \Tr_{\ff} \overline{U}^n \right] \ ,
\ee
where $\ff$ is the fundamental representation of U$(\infty)$ and\footnote{By ${\rm U}(\infty)$, we simply mean that we evaluate every expression for ${\rm U}(N)$ with finite $N$ and then take the large $N$ limit. Note that terms that scale like $q^N$ vanish in this limit, because ${\rm Im}(\tau) > 0$.} 
\be\label{zdefgen}
z_i = q^{i+h-1}  \  , \qquad \Tr_{\ff} U^n = \frac{q^{nh}}{1-q^n} \ .
\ee
This can be re-expressed using the Frobenius relation
\be
\Tr_{\ff} U^{j_1}\, \Tr_{\ff} U^{j_2} \, \cdots \Tr_{\ff} U^{j_m} = \sum_R \chi^{(S)}_R(j_1,\dots,j_m)\Tr_R U \ ,
\ee
where $\chi^{(S)}$ is a character of the permutation group and the sum is over Young tableaux with $\sum j_i$ boxes. Next we expand the exponential,
\bea
Z_{\rm scal}(h) &=& \prod_n \sum_{k_n} {1\over k_n ! \, n^{k_n}} (\Tr_f U^n)^{k_n} (\Tr_f \overline{U}^n)^{k_n}\\
&=& \sum_{\vec{k}} \prod_n {1\over k_n! \, n^{k_n}}(\Tr_f U^n)^{k_n} (\Tr_f \overline{U}^n)^{k_n}\\
&=& \sum_{\vec{k}} {1\over z_{\vec{k}}}\sum_{R, \overline{R}}\chi_R^{(S)}(\vec{k}) \,
\chi_{\overline{R}}^{(S)}(\vec{k})\ \Tr_{R} U \  \Tr_{\overline{R}} \overline{U} \ ,
\eea 
where 
\be
z_{\vec{k}} = \prod_n k_n! \, n^{k_n} \ .
\ee
Here $\vec{k}$ labels a conjugacy class of the permutation group $S_\ell$ on 
$\ell = \sum_{n} n k_n$ elements; the class has $k_1$ cycles of length 1, 
$k_2$ cycles of length 2, {\it etc.} The orthogonality property of the characters is
\be
\sum_{\vec{k}} |C(\vec{k})| \, \chi_{R}^{(S)}(\vec{k}) \chi_{\overline{R}}^{(S)}(\vec{k}) 
= \ell\, !\,\, \delta_{R \overline{R}} \ ,
\ee
where $|C(\vec{k})|$ is the number of elements in the conjugacy class $\vec{k}$, explicitly
given as
\be
|C(\vec{k})| = \frac{\ell\, !}{z_{\vec{k}}}   \ .
\ee
Therefore we have finally
\be\label{realscalarres}
Z_{\rm scal}(h) = \sum_R \Tr_{R} U \ \Tr_{R} \overline{U} \ .
\ee
By definition, $\Tr_{R} U$ is the character of the representation specified by the
Young tableau $R$ evaluated with chemical potentials $\log{z_i}$, where $z_i$ is
defined in (\ref{zdefgen}). 
For the case at hand, $h=h_\pm$ see eq.~(\ref{hpm}), this simplifies in the large $N$ limit to 
\be
\Tr_{R}(U) = \chiu_R(q^{i+h_\pm-1}) =  q^{\pm \frac{\lambda}{2} B(R)} \, \chiu_R(q^{i-\frac{1}{2}} ) 
= P_R^\pm(q) \ , 
\ee
where the second identity follows directly from (\ref{schurchar}), and we have used the notation 
introduced in (\ref{schurpm}) in the last step. Collection all the contributions to (\ref{bulkz}) 
we can now rearrange the gravity partition function into its final form,
\be\label{gravityfinal}
Z_{\rm bulk} = (q \bar{q})^{-c/24}Z_{\rm hs} 
\sum_{R_{+}, R_{-}, S_{+}, S_{-}}|P_{R_+}^+(q) P_{S_+}^+(q) P_{R_-}^-(q) P_{S_-}^-(q)|^2 \ .
\ee

\section{CFT Partition Function}\label{s:cft}

In this section we derive explicit expressions for the CFT characters in the 't~Hooft limit. First
we study the behaviour of the branching functions in this limit. Then we shall explain which terms
actually survive once we remove the contributions from the states that decouple. A more
detailed explanation of why these states decouple will be given in Section~\ref{s:fusion}.

\subsection{The Large $N$, $k$ Limit of Branching functions}

Recall that we are considering the coset CFT (\ref{gencos}), 
whose representations are labelled by $(\Lambda_+;\Lambda_-)$, with $\Lambda_+$ 
and $\Lambda_-$ being integrable highest weights of $\mathfrak{su}(N)_k$ and $\mathfrak{su}(N)_{k+1}$, 
respectively. The character of the coset representation $(\Lambda_+;\Lambda_-)$ can be expressed in
terms of a branching function (see eq.\ (7.51) of \cite{Bouwknegt:1992wg})
\begin{equation}\label{brfn}
b_{(\Lambda_+;\Lambda_-)}(q) =
 {1 \over \eta(q)^{N-1}} \sum_{w \in \hat{W}} 
 \epsilon(w) q^{{1 \over 2 p(p+1)} ( (p+1)w(\Lambda_+ + \rho) - p (\Lambda_- + \rho) )^2}\ .
\end{equation}
Here $p=k+N$, and the sum is over the full affine Weyl group. The affine Weyl group is the 
semidirect product of the finite Weyl group and translations by elements ${\bf P}$ of the root lattice, 
and its action is given by
\be
w(\Lambda + \rho) = w_{\rm finite} (\Lambda+\rho) + (k+N) {\bf P} \ ,
\ee
where we think of $\Lambda$ and $\rho$ as a weight and the Weyl vector 
of the finite dimensional Lie algebra, respectively (see also \cite{Bouwknegt:1991gf}, after 
eq.\ (3.9)). In the  't~Hooft limit, 
\begin{equation}\label{thft}
N,k \rightarrow \infty \qquad \hbox{with} \qquad \lambda=\frac{N}{N+k} \quad {\rm fixed,}
\end{equation}
eq.\ (\ref{brfn}) simplifies considerably. First, we can drop the sum over the affine translations, as the 
corresponding terms will give rise to terms of order $q^{k+N}$, which we can ignore in the 't~Hooft limit
(see also \cite{Bouwknegt:1992wg}, below eq.\ (7.40)); thus
we can restrict the sum to the finite Weyl reflections. Furthermore, since finite Weyl reflections leave the 
norm of a vector unchanged, the exponent of $q$ in (\ref{brfn}) can be written as 
\begin{equation}\label{expon3}
{2p(p+1)+1\over 2 p(p+1)}\rho^2 +\bigl(1+\tfrac{1}{p}\bigr)\, C_2(\Lambda_+)
+ \bigl(1- \tfrac{1}{p+1}\bigr)\, C_2(\Lambda_-) 
-\, \langle w(\Lambda_++\rho), (\Lambda_- + \rho) \rangle \ , 
\end{equation}
where we have used that the quadratic Casimir equals
\begin{equation}
C_2(\Lambda) = \tfrac{1}{2} \bigl( \Lambda^2 + 2 \langle \Lambda, \rho \rangle \bigr) \ .
\end{equation}
Let us write the representation $\Lambda_\pm$ as $\Lambda_\pm = (\overline{R}_{\pm},S_\pm)$, see
figure~\ref{fig:tableaux}. It was shown in eq.\ (2.7) of \cite{Gross:1993hu} that 
\be\label{C2sum}
C_2(\Lambda_\pm) = C_2(R_\pm) + C_2(S_\pm) + {\cal O}\bigl(\tfrac{1}{N}\bigr) \ .
\ee
The representations $R_\pm$ and $S_\pm$ are described by Young tableaux with finitely
many boxes. For such a representation $L$, let us denote by $r_i$ the number of boxes in the 
$i^{\rm th}$ row, while $c_j$ is the number of boxes in the $j^{\rm th}$ column, so that
\begin{equation}
B(L) = \sum_i r_i = \sum_j c_j  
\end{equation}
is the total number of boxes of the Young tableau corresponding to $L$. Defining
\begin{equation} 
D(L) = \sum_i r_i^2 - \sum_j c_j^2 \ ,  
\end{equation}
it then follows from \cite[eq.\ (2.1)] {Gross:1993hu} (see also appendix \ref{a:orthobasis}) 
that the quadratic Casimir has the expansion
\begin{equation}\label{casexp}
C_2(L) = \tfrac{1}{2} B(L) N  + \tfrac{1}{2} D(L) + {\cal O}\bigl(\tfrac{1}{N}\bigr)  \ .
\end{equation}
Hence we find for the quadratic Casimir of $\Lambda_\pm$
\be
C_2(\Lambda_\pm) = \tfrac{1}{2} B_\pm\, N + \tfrac{1}{2} D_\pm +  {\cal O}\bigl(\tfrac{1}{N}\bigr)  \ ,
\ee
where $B_\pm = B(R_\pm) + B(S_\pm)$, see eq.\ (\ref{Bpm}),
and $D_\pm = D(R_\pm) + D(S_\pm)$. In the 't~Hooft limit we have 
\begin{equation}
\frac{1}{p} = \frac{1}{p+1} = \frac{\lambda}{N} 
\end{equation}
and hence (\ref{brfn})  simplifies to 
\begin{equation}\label{brfnsimp}
b_{(\Lambda_+;\Lambda_-)}(q) \cong q^{\frac{(N-1)}{24} \lambda^2}
 \frac{q^{C_2(\Lambda_+) + C_2(\Lambda_-) }\, 
q^{\frac{\lambda}{2} (B_+-B_-)} }{\eta(q)^{N-1}}  \, 
q^{\rho^2}\,  
\sum_{w \in W} \epsilon(w) q^{-\langle w(\Lambda_+ + \rho), \, (\Lambda_- + \rho)\rangle }\ ,
\end{equation}
where $\cong$ denotes identities that are true up to terms that go to zero as $N\rightarrow \infty$.
We have also used that for $\mathfrak{su}(N)$, $\rho^2 =\frac{N(N^2-1)}{12}$ , and hence 
we get a term proportional to $\tfrac{1}{24}(N-1) \lambda^2$ from the first term of (\ref{expon3}).
The Weyl denominator formula for $\mathfrak{su}(N)$ states 
that
\begin{equation}\label{weylde}
\sum_{w \in W} \epsilon(w) q^{-\langle w(\rho), \rho\rangle} = 
q^{-\rho^2} \prod_{n=1}^{N-1} (1-q^n)^{N-n} \ .
\end{equation}
Solving for $q^{\rho^2}$ and plugging into (\ref{brfnsimp}) then leads to 
\be
\label{brfnsimp1}
b_{(\Lambda_+;\Lambda_-)}(q) \cong
q^{-\frac{c}{24}}\, 
q^{\frac{\lambda}{2} (B_+-B_-)} \, q^{C_2(\Lambda_+) + C_2(\Lambda_-) }\, 
\frac{\sum_{w \in W} \epsilon(w)\, q^{-\langle w(\Lambda_+ + \rho), \, (\Lambda_- + \rho)\rangle }}
{\sum_{w \in W} \epsilon(w) \, q^{-\langle w(\rho), \rho\rangle} }\, \tilde{M}(q) \ ,
\ee
where we have used that $c = (N-1) (1-\lambda^2)$ and $\tilde{M}(q)$ is as defined in (\ref{macdef}).
Next we observe that the above ratio can be expressed in terms of $S$-matrix elements 
of the affine algebra as (see {\it e.g.}\ \cite[eq.\ (2.7.24)]{Fuchs:1992nq})
\be
\frac{\sum_{w \in W} \epsilon(w)\, q^{-\langle w(\Lambda_+ + \rho), \, (\Lambda_- + \rho)\rangle }}
{\sum_{w \in W} \epsilon(w) \, q^{-\langle w(\rho), \rho\rangle} }\, = \frac{S_{\Lambda_+\Lambda_-}}{S_{00}} \ ,
\ee
while the exponential of the Casimir is related to the modular $T$-matrix,
\be
q^{C_2(\Lambda)} = \frac{T_{\Lambda\Lambda}}{T_{00}} \ .
\ee
Both of these identities hold for $q=\exp(\tfrac{2\pi i}{k+N})$. Thus we can rewrite 
\be
q^{C_2(\Lambda_+) + C_2(\Lambda_-)}\, 
\frac{\sum_{w \in W} \epsilon(w)\, q^{-\langle w(\Lambda_+ + \rho), \, (\Lambda_- + \rho)\rangle }}
{\sum_{w \in W} \epsilon(w) \, q^{-\langle w(\rho), \rho\rangle} } = \frac{1}{S_{00} T_{00}^2} \, 
T_{\Lambda_+\Lambda_+}\, S_{\Lambda_+\Lambda_-}\, 
T_{\Lambda_-\Lambda_-} \ .
\ee
Following \cite{Aganagic:2002qg} we have the identity 
\be
T_{0,0}^{-1} \, T_{\Lambda_+\Lambda_+}\, S_{\Lambda_+\Lambda_-} \, T_{\Lambda_-\Lambda_-}
= \sum_\Lambda N_{\Lambda_+\overline{\Lambda}_-}^{\ \ \ \ \ \ \Lambda}\ T_{\Lambda\Lambda} S_{\Lambda 0} \ ,
\ee
where $N_{\Lambda_+ \overline{\Lambda}_-}{}^{\Lambda}$ are the fusion rules, and 
$\overline{\Lambda}_{-}$ is the conjugate representation to $\Lambda_{-}$; this  can be proved
using Verlinde's formula
\bea
\sum_\Lambda N_{\Lambda_+ \overline{\Lambda}_-}{}^{\Lambda}\ T_{\Lambda\Lambda} S_{\Lambda0} &=& 
\sum_{\Lambda,\Pi} \frac{S_{\Lambda_+\Pi}\, S_{\overline{\Lambda}_- \Pi} S^{-1}_{\Pi\Lambda}}{S_{0\Pi}} \, 
T_{\Lambda\Lambda} S_{\Lambda 0}^{-1} = \sum_{\Pi} \frac{ S_{\Lambda_+\Pi}S_{\overline{\Lambda}_-\Pi} }
{ S_{0\Pi}} (S^{-1} T S^{-1})_{\Pi 0} \notag\\
&=& \sum_{\Pi} \frac{ S_{\Lambda_+\Pi}S_{\overline{\Lambda}_- \Pi} }{ S_{0\Pi}} 
(T^{-1} S T^{-1})_{\Pi 0} = T_{00}^{-1} (S T^{-1} S)_{\Lambda_+ \overline{\Lambda}_-} \notag\\
&=& T_{00}^{-1} \, (T S^{-1} T)_{\Lambda_+ \overline{\Lambda}_-} 
= T_{00}^{-1} \, T_{\Lambda_+ \Lambda_+} S_{\Lambda_+ \Lambda_-} T_{\Lambda_-\Lambda_-} \ .\notag
\eea
Here we have used the identity $(ST)^3=S^2=C$ twice, as well as the fact that $S$ is symmetric and 
$S^{-1}_{\Lambda 0} = S_{\Lambda^\ast 0} =  S_{\Lambda 0}$.
Putting everything together we then conclude that (\ref{brfnsimp1}) agrees with 
\be
\label{brfnsimp2}
b_{(\Lambda_+;\Lambda_-)}(q) \cong
q^{-\frac{c}{24}}\,  \tilde{M}(q)\, 
q^{\frac{\lambda}{2} (B_+-B_-)} \, 
\sum_{\Lambda} N_{\Lambda_+ \overline{\Lambda}_-}{}^{\Lambda} \, 
q^{C_2(\Lambda)} \frac{S_{\Lambda 0}}{S_{00}} \,  \ .
\ee
Note that the ratio of $S$-matrix elements that appears in (\ref{brfnsimp2}) actually
equals the quantum dimension of $\Lambda$, 
\be\label{qdim}
 \frac{S_{\Lambda 0}}{S_{00}}  = \dim_q(\Lambda) =
\frac{\sum_{w \in W} \epsilon(w)\, q^{-\langle w(\Lambda+\rho), \, \rho\rangle }}
{\sum_{w \in W} \epsilon(w) \, q^{-\langle w(\rho), \rho\rangle} } \ , 
\ee
which appears in the branching function for $(\Lambda;0)$; thus we can, in
particular, rewrite the right-hand-side of (\ref{brfnsimp2}) as 
\be\label{generalbranch}
b_{(\Lambda_+;\Lambda_-)}(q)  \cong q^{\frac{\lambda}{2} (B_+-B_-)}
\sum_{\Lambda} N_{\Lambda_+\overline{\Lambda}_-}{}^{\Lambda} \,  
q^{- \frac{\lambda}{2} B(\Lambda)} \, b_{(\Lambda;0)}(q) \ ,
\ee
thus reproducing (\ref{generalbranch1}). 
Note that the fusion rule coefficients that appear in (\ref{generalbranch}) stabilise for sufficiently
large $k$ and $N$. Both sides can be thought of as power series in $q$, and the previous
argument shows that the low order terms of the two power series ({\it i.e.}\ the terms whose 
order is less than $k$ or $N$)  agree for all $q$ of the form $q=\exp(\tfrac{2\pi i}{k+N})$
with $k$ and $N$ sufficiently large. This is then sufficient to prove that (\ref{generalbranch})
defines an identity of power series for $N,k\rightarrow \infty$.

\subsection{Relating the Branching Functions to $\mathfrak{u}(N)$ Characters}

In the previous subsection we have reduced the branching function of $(\Lambda_+;\Lambda_-)$ to 
that of $(\Lambda;0)$. Using the formula for the quantum dimension (\ref{qdim}), we can rewrite
the expression appearing in (\ref{generalbranch}) as 
\be
q^{-\frac{\lambda}{2} B(\Lambda)}\, b_{(\Lambda;0)}(q) \cong  q^{-\frac{c}{24}}\, 
 \tilde{M}(q) \,  q^{C_2(\Lambda)} \,  \dim_q(\Lambda)  \ . 
\ee
Next we want to show that $q^{C_2(\Lambda)} \,  \dim_q(\Lambda)$ can be interpreted as a 
character for ${\mathfrak u}(N)$. First, it follows from \cite[eq.\ (B.1)]{Aganagic:2004js} and (\ref{C2sum}) that 
for $\Lambda=(\overline{R},S)$ the quantum dimension factorizes as 
\be
q^{C_2(\Lambda)} \,  \dim_q(\Lambda) \cong q^{C_2(R)} \dim_q(R) \cdot q^{C_2(S)}\, \dim_q(S) \ .
\ee
Thus it is sufficient to consider the case of a representation $L$ with finitely many boxes, for which 
\be
\dim_q(L) = \chi_L^{\mathfrak{su}(N)} (\tilde{z}_i) \ , \qquad
\tilde{z}_i = q^{i-\frac{N+1}{2}} \ ,
\ee
since, in the orthogonal basis, the Weyl vector has components $\rho_i = \tfrac{N+1}{2}-i$.
Note that the product of the $\tilde{z}_i$ is unity, $\prod_{i=1}^{N} \tilde{z}_i =1$. The  
$\mathfrak{su}(N)$ characters are essentially equal to the ${\mathfrak u}(N)$ characters
given in (\ref{schurchar}), except that for the latter one does not impose that the product of the 
chemical potentials $z_i$ equals unity. Writing 
\be
\tilde{z}_i = q^{-\frac{N}{2}}\, z_i \ , \qquad z_i = q^{i - \frac{1}{2}} \ ,
\ee
and using (\ref{schurchar}), we then find
\be\label{1.30}
\dim_q(L) = \chi_L^{\mathfrak{su}(N)} (\tilde{z}_i) =
\chi_L^{{\mathfrak u}(N)} (\tilde{z}_i)  = q^{-\frac{N}{2} B(L)} \, \chi_L^{{\mathfrak u}(N)} (z_i) \ , 
\ee
where  $\sum_j r_j = B(L)$ is the total number of boxes. Together with the 
expansion of the quadratic Casimir (\ref{casexp}), we thus conclude that
\be\label{1.31}
q^{C_2(L)} \,  \dim_q(L) \cong q^{\frac{1}{2}\, D(L)} \,
q^{\frac{NB(L)}{2}} \, q^{-\frac{N}{2} B(L)} \, \chi_L^{{\mathfrak u}(N)} (z_i) 
\cong q^{\frac{1}{2}\, D(L)} \,\chi_L^{{\mathfrak u}(N)} (z_i)  \ . 
\ee
Finally, one shows (see appendix \ref{a:orthobasis})
\be\label{uNflip}
q^{\frac{1}{2}\, D(L)} \,\chi_L^{{\mathfrak u}(N)} (z_i) 
= \chi_{L^T}^{{\mathfrak u}(N)} (z_i)  = P_{L^T}(q) \ ,
\ee
where $L^T$ is the representation whose Young tableau has been flipped relative to $L$, and
we have used the notation introduced in eq.~(\ref{schurpm}). Putting everything together it then follows 
that
\be
q^{-\frac{\lambda}{2} B(\Lambda)}\, b_{(\Lambda;0)}(q) \cong q^{-\frac{c}{24}}\,    \tilde{M}(q)  \, 
P_{R^T}(q) \, P_{S^T}(q)  \ ,
\ee
and thus the general formula is 
\be\label{branchfin}
b_{(\Lambda_+;\Lambda_-)}(q)  \cong q^{\frac{\lambda}{2} (B_+ - B_-) }\, 
q^{-\frac{c}{24}}\,    \tilde{M}(q)\,
\sum_{\Lambda=(\overline{R}, S)} N_{\Lambda_+ \overline{\Lambda}_-}{}^{\Lambda} \,  
P_{R^T}(q) \, P_{S^T}(q)  \ . 
\ee
At large $k$, the fusion coefficients $N_{\Lambda_+ \overline{\Lambda}_-}{}^{\Lambda}$ are simply the
Clebsch-Gordon series for the tensor product decomposition of the corresponding 
$\mathfrak{su}(N)$ representations.

\subsection{Removing the Decoupled States}

The Clebsch-Gordon series for $\Lambda_+ \otimes \overline{\Lambda}_{-}$ contains
only representations $\Lambda$ for which 
\be
B(\Lambda) \leq B(\Lambda_+) + B(\Lambda_-)\ .
\ee
Here $B(\Lambda) = B(R) + B(S)$ is the
sum of boxes and `antiboxes', and similarly for $B(\Lambda_+)$ and $B(\Lambda_-)$. 
In general $B(\Lambda)$ may however be strictly smaller than $B(\Lambda_+)+B(\Lambda_-)$;
in particular, this will be the case if a box from $\Lambda_+$ cancels against an antibox
from $\overline{\Lambda}_{-}$, or an antibox from $\Lambda_+$ cancels against a box from
$\overline{\Lambda}_{-}$.

As we have mentioned before and as will be explained in more detail in the following
section~\ref{s:fusion}, the representation $(\Lambda_{+};\Lambda_{-})$ is typically
indecomposable, and in the amplitudes of the limit theory only a subspace of states survive.
Our analysis suggests that these states are precisely associated to those 
$\Lambda$ where no boxes or antiboxes have cancelled. Thus for the calculation
of the actual CFT partition function we need to restrict the sum in (\ref{branchfin}) to those $\Lambda$,
for which $B(\Lambda) = B(\Lambda_+) + B(\Lambda_{-})$.

In terms of the description of the representations $\Lambda_\pm$ in terms of finite Young tableaux
$\Lambda_\pm = (\overline{R}_\pm,S_\pm)$, see figure~\ref{fig:tableaux},
the condition that no boxes or antiboxes cancel simply means that the resulting $\Lambda$ will be 
contained in 
\be
\Lambda \in (\overline{R_{+} \otimes S_{-}} , S_+ \otimes R_-) \ .
\ee
Thus the actual character of the CFT representation is 
\be\label{branchfin1}
{\rm ch}_{(\Lambda_+;\Lambda_-)}(q)  \cong q^{\frac{\lambda}{2} (B_+ - B_-) }\, 
q^{-\frac{c}{24}}\,    \tilde{M}(q)\,
\sum_{R, S} N_{R_+ S_-}{}^{R} \,   N_{R_{-} S_{+}}{}^{S}\, P_{R^T}(q) \, P_{S^T}(q)  \ ,
\ee
where now $N_{R_\pm S_{\mp}}{}^{L}$ describe the Clebsch-Gordon decomposition of
finite Young tableaux. Since the Schur functions are just characters of ${\mathfrak u}(N)$
representations it follows that
\be
\sum_{R} N_{R_+ S_-}{}^{R} P_{R^T}(q) = P_{R_{+}^T}(q) \, P_{S_{-}^T}(q) \ ,
\ee
and similarly for the term involving $P_{S^T}(q)$. Hence we arrive at 
\begin{eqnarray}\label{branchfin2}
{\rm ch}_{(\Lambda_+;\Lambda_-)}(q)  &  \cong & q^{\frac{\lambda}{2} (B_+ - B_-) }\, 
q^{-\frac{c}{24}}\,    \tilde{M}(q)\,
P_{R_{+}^T}(q) \, P_{S_{-}^T}(q)  \, P_{R_{-}^T}(q) P_{S_{+}^T}(q)  \notag \\
& = & q^{-\frac{c}{24}}\,    \tilde{M}(q)\,
P^+_{R_{+}^T}(q) \, P^+_{S_{+}^T}(q)\,     \, P^-_{R_{-}^T}(q) \, P^-_{S_{-}^T}(q) \ , 
\end{eqnarray}
where in the last line we have absorbed the $\lambda$-dependent prefactor into the definition of 
$P^\pm_L(q)$, see eq.~(\ref{schurpm}). This then reproduces precisely
eq.~(\ref{fcft}).


\section{Decoupling in the Large-N CFT}\label{s:fusion}

In this section we give evidence for the claims of section~\ref{s:cft}
regarding the structure of the li\-mi\-ting representations and their behaviour in 
correlation functions. We shall first explain why 
the indecomposability of the representation, for example for the representation depicted in
(\ref{ffdia}), leads to a 
decoupling of the subrepresentations from correlation functions. Then we shall
explain why the representations are, in fact, indecomposable. After explaining the general
strategy of our calculation in section~\ref{s:genstra}, we demonstrate the indecomposability 
for the case of $({\rm f};{\rm f})$ in section~\ref{s:ffde}. We have also tested a number of 
other cases, and the details of the corresponding calculations are described in the appendix, 
see also section~\ref{ss:othercases}.

\subsection{Decoupling of Null States}\label{ss:decouplingb}

In the following we want to explain why the indecomposability of the representation 
leads to a decoupling of the subrepresentations from correlation functions. For
concreteness we shall concentrate on the case $({\rm f};{\rm f})$, for which the 
structure of the resulting representation was already described in section~\ref{s:sumCFT}, 
see (\ref{ffdia}); as we shall see, the arguments for the other cases are essentially
identical. 

The argument is more or less the same as for usual null vectors, except that the 
cyclic state, {\it i.e.}\ the state from which any state can be obtained by the action of the modes, 
is the state $\psi$ at $h=1$, and the `null vector' is the vector $\omega$ at $h=0$, see
(\ref{ffdia}). However, 
for the usual decoupling argument it is not actually relevant whether the null vector has higher 
or smaller conformal weight than the cyclic vector, and thus the argument goes through essentially 
unaltered. 

To understand this in more detail, recall that it follows from the usual factorisation arguments that a 
state $\chi$ will only be non-zero
in any correlation function provided it is non-zero in a suitable 2-point function. We are interested in
the correlation functions involving an arbitrary number of the fundamental fields $(0;{\rm f})$,
$(0;\bar{\rm f})$, $({\rm f};0)$, and $(\bar{\rm f};0)$. Thus we only need to consider two point functions
where both states live in representations that appear in multiple fusion products of these representations. 
Based on the structure of the fusion rules for finite $N$, the only non-zero 2-point function involving
$({\rm f};{\rm f})$ is then the one where the other state transforms in the $(\bar{\rm f};\bar{\rm f})$ 
representation, for which the fusion analysis is essentially identical. 

Following the notation of (\ref{ffdia}) we denote the 
`highest weight' states of $({\rm f};{\rm f})$  by $\omega$  and 
$\psi$; similarly, we call the two `highest weight' states of $(\bar{\rm f};\bar{\rm f})$ $\bar\omega$  and 
$\bar\psi$. Since the two-point functions are diagonal with respect to conformal weight, the only 
potentially non-zero two-point functions are then 
\be
\langle \bar\omega | \omega \rangle  \ , \qquad \hbox{and} \qquad
\langle \bar\psi | \psi \rangle  \ .
\ee
Note in particular, that there is no other state at $h=1$ in $({\rm f};{\rm f})$ with which
$\bar\psi$ could have a non-trivial 2-point function, and similarly for $\psi$. Now if 
$\langle \bar\psi | \psi \rangle=0$, then $\psi$ will be zero in all correlation functions.
But then the same will be true for any descendant of $\psi$, {\it i.e.}\ any state that can be
obtained from $\psi$ by the action of modes. Thus the whole $({\rm f};{\rm f})$ representation
and the whole $(\bar{\rm f};\bar{\rm f})$ representation would be zero in all correlation functions. 
While we cannot a priori exclude this possibility, this is certainly not what we should expect. 

So let us then assume that $\langle \bar\psi | \psi \rangle\neq 0$, say $\langle \bar\psi | \psi \rangle=1$. 
Then, given that $\bar\omega = L_1 \bar\psi$, we conclude
\be\label{central}
\langle \bar\omega | \omega \rangle  = \langle L_1 \bar\psi | \omega \rangle 
= \langle \bar\psi | L_{-1} \omega \rangle = 0  \ ,
\ee
since $L_{-1} \omega\neq \psi$ --- in fact $L_{-1}\omega=0$. Thus it follows that 
$\langle \bar\omega | \omega \rangle=0$, and hence $\omega$, as well as the whole subrepresentation
generated from it, will vanish in all correlation functions.
In the more general cases we have studied, while there are potentially other states at the same level as the cyclic state, these always turn out to have vanishing overlap since they carry different higher spin charge.

\subsection{The General Strategy of the Calculation}\label{s:genstra}

Next we want to explain why the limiting representations are indeed indecomposable.  Recall from 
section~\ref{s:cft} that we expect this phenomenon to arise for representations of the form
$(\Lambda_+, \Lambda_-)$, where $\Lambda_+$ and $\Lambda_-$ have common boxes or common antiboxes. We can think of these
representations as being defined by the fusion of $(\Lambda_+,\Lambda_-)=(\Lambda_+,0)\otimes (0,\Lambda_-)$. Thus we need to
understand the structure of the fusion product of $(\Lambda_+,0)$ and $(0,\Lambda_-)$. 
A technology to study 
general fusion products (without assuming anything about their structure) was proposed some time
ago by  Nahm \cite{Nahm:1994by} and then further developed in \cite{Gaberdiel:1996kx}. It has been
successfully applied to logarithmic conformal field theories where this approach is one of the standard
methods by now. 

The basic idea is to think of the fusion of two representations as the tensor product 
${\cal H}_1 \otimes {\cal H}_2$ of the corresponding
representation spaces; this tensor product space carries an action of the symmetry algebra of the conformal
field theory ({\it i.e.}\ the limit of the $\W_N$ algebras in our context). The fusion rules are then the 
`Clebsch-Gordon'  coefficients describing the decomposition of the tensor product in terms
of irreducible (or in general indecomposable) representations. 

The main problem with this idea is that all representation spaces involved are infinite dimensional.
Thus in order to actually turn this into a computationally feasible algorithm, we need to cut the problem
down to size. For example, we could try to concentrate in a first step on the highest weight
states of the fusion product. The highest weight states are characterised by the property that they are
annihilated by all positive modes. Unfortunately, this is a condition that is very hard to implement in 
the above setting, but there is an almost equivalent alternative: at least in the standard highest weight
case, the highest weight 
states are not only annihilated by all positive modes, but they also have the property that they cannot be 
obtained from any other state by the action of the negative modes. Thus we may think of the `highest weight'
states as being described by the quotient space
\be
({\cal H}_1 \otimes {\cal H}_2 )^{(0)} = ({\cal H}_1 \otimes {\cal H}_2 ) / {\cal A}_{<0} ({\cal H}_1 \otimes {\cal H}_2 ) \ ,
\ee
where ${\cal A}_{<}$ is the algebra of negative modes, and we quotient out by all states that can be obtained 
by the action of a negative mode on any state in $({\cal H}_1 \otimes {\cal H}_2 )$. Provided that the
fusion rules are finite, this quotient space is then finite-dimensional, and we can calculate its dimension,
the action of the zero modes on it, {\it etc}. 

Up to now we have not gained much relative to the usual treatment of fusion, but it should now be clear
how we can proceed. In addition to the `highest weight' space $({\cal H}_1 \otimes {\cal H}_2 )^{(0)}$ we
can also consider larger quotient spaces, where we divide out by smaller and smaller subalgebras.
In particular, we can define 
\be\label{lev1}
({\cal H}_1 \otimes {\cal H}_2 )^{(1)} = ({\cal H}_1 \otimes {\cal H}_2 ) / 
{\cal A}_{<-1} ({\cal H}_1 \otimes {\cal H}_2 ) \ ,
\ee
where we now only divide out states that can be obtained by the action of modes whose total mode 
number is less than $-1$, such as 
$L_{-2}$, $L_{-3}$, $W^{(3)}_{-2}$, $\ldots$, as well as $L_{-1} L_{-1}$, $L_{-1} W^{(3)}_{-1}$, $\ldots$ {\it etc.}
Provided that the fusion rules are finite, $({\cal H}_1 \otimes {\cal H}_2 )^{(1)}$ will again be 
a finite-dimensional vector space on which we can calculate the action of the zero modes, {\it etc}. 
In addition, we can now, however, also determine the action of some of the positive modes. In particular,
any $+1$ mode such as $L_1$, $W^{(3)}_1$, {\it etc.} defines a well-defined map
\be
L_1 : ({\cal H}_1 \otimes {\cal H}_2 )^{(1)} \rightarrow ({\cal H}_1 \otimes {\cal H}_2 )^{(0)} \ ,
\ee
and we can thus determine its action, at least on those states that are visible in 
$({\cal H}_1 \otimes {\cal H}_2 )^{(1)}$.

It should be clear that we can continue in this way: by determining larger  quotient spaces
(where we divide out smaller  subspaces of $({\cal H}_1 \otimes {\cal H}_2 )$) we can unravel
the structure of $({\cal H}_1 \otimes {\cal H}_2 )$ more  precisely, since we get access to the action
of higher  modes as we proceed. Obviously, the analysis will get harder and harder (and it 
is not feasible to do it for all such quotient spaces in closed form), but it is often enough to determine the
first few such quotient spaces in order to deduce the structure of the resulting representation. For example, 
for the case at hand, the fusion of $({\rm f};0)$ and $(0;{\rm f})$, it will be enough to determine (\ref{lev1}) 
since then we can determine how $L_1$ will act on the potential primary state at $h=1$. In particular,
this will allow us to detect the presence (or absence) of the arrow from $\psi$ to $\omega$ in (\ref{ffdia}).

\subsection{The Calculation for $({\rm f};0) \otimes (0;{\rm f})$}\label{s:ffde}

Let us explain this general analysis with the simplest example; we shall comment on the other cases
we have studied in section~\ref{ss:othercases}. This corresponds to both 
$R$ and $S$ being the fundamental representation\footnote{The same example was considered in \cite{Gaberdiel:2010pz}, 
where it was suggested that the appearance of a new null vector in the large $N$ limit could lead to the 
decoupling of problematic states.  This suggestion, based on a similar phenomenon in the Virasoro minimal 
models \cite{Roggenkamp:2003qp}, turns out to be correct but the details are more intricate  than the 
Virasoro case (where the representations are completely reducible).}. 
Before taking the 't~Hooft limit $N\rightarrow\infty$ 
the fusion of $({\rm f};0)$ and $(0;{\rm f})$ is an irreducible representation with
\be
h({\rm f};{\rm f}) = \frac{N^2-1}{2 N (k+N)(k+N+1)} \ .
\ee
In the 't Hooft limit we obviously have $h({\rm f};{\rm f})=0$, and hence $L_{-1}({\rm f};{\rm f})$ will
become a null vector. Thus we may expect that, in the 't Hooft limit, the representation $({\rm f};{\rm f})$ will
become a direct sum of representations. In fact, this is also suggested by the character of $({\rm f};{\rm f})$
which becomes in the 't~Hooft limit
\be\label{deco}
\chi_{({\rm f};{\rm f})} = \chi_{(0;0)} + \frac{q^1}{(1-q)^2} \prod_{s=2}^{\infty} \prod_{n=s}^{\infty} \frac{1}{(1-q^n)} \ ,
\ee
indicating that $({\rm f};{\rm f})$ splits up as the direct sum of the vacuum representation $(0;0)$, and a
second representation with highest weight $h=1$, whose character is the second term in (\ref{deco}). 
In fact, this is precisely what happens for the $c\rightarrow 1$ limit of the minimal models, as argued
in \cite{Roggenkamp:2003qp}, see also \cite{Graham:2001tg}. 

However, it is clear that this cannot quite happen in our case. For $\lambda\neq 0$, the
two representations $({\rm f};0)$ and $(0;{\rm f})$ have different conformal dimensions, 
$h=h_\pm =\tfrac{1}{2}(1\pm \lambda)$, and the fusion of $({\rm f};0)$ and $(0;{\rm f})$ therefore cannot
contain the actual vacuum. (This argument does not apply in the $c\rightarrow 1$ limit of the minimal models, 
since there the relevant two representations, namely $(2,1)$ and $(1,2)$ both have conformal dimension
$h=\tfrac{1}{4}$ in the limit.) In order to find out what precisely happens we can perform the fusion
analysis that was outlined above. 

First we need to determine the low-lying null vector 
relations that characterise these two representations. We choose the normalisation of  
$W\equiv W^{(3)}$ and  $U\equiv W^{(4)}$ so that their eigenvalues $h=L_0$, $w=W_0$ and $u=U_0$ 
on the ground states are 
\be
\phi_1 \equiv ({\rm f};0) : \quad 
h_1 = \tfrac{1}{2} (1+\lambda) \ , \qquad w_1 = - (1+\lambda)(2+\lambda) \ , \qquad
u_1 = (1+\lambda)(2+\lambda)(3+\lambda) \ ,
\ee
and 
\be
\phi_2 \equiv (0;{\rm f}) : \quad 
h_2 = \tfrac{1}{2} (1-\lambda) \ , \qquad w_2 =  (1-\lambda)(2-\lambda)  \ , \qquad
u_2 = (1-\lambda)(2-\lambda)(3-\lambda) \ .
\ee
Both of these representations have the property that they only have one descendant state at conformal 
dimenison one above the ground state, as follows directly from their characters
\be
\chi_{({\rm f};0)}(q) = \frac{q^{\frac{1}{2}(1+\lambda)}}{(1-q)} \, 
\prod_{s=2}^{\infty}\prod_{n=s}^{\infty} \frac{1}{(1-q^n)} = 
q^{\frac{1}{2}(1+\lambda)} \Bigl( 1 + q + 2 q^2 + \cdots \Bigr) \ ,
\ee
and similarly for $\chi_{(0;{\rm f})}(q)$. Thus both $W_{-1}\phi_j$ and $U_{-1}\phi_j$  must be proportional to 
$L_{-1}\phi_j$; the proportionality constants can be worked out using the condition that the difference must
be annihilated by $L_1$, thus leading to 
\be
\begin{array}{rclrcl}
W_{-1} \phi_1 &= &  - 3 (2+\lambda) L_{-1} \phi_1  \qquad 
& W_{-1} \phi_2 &= & 3 (2-\lambda)   L_{-1} \phi_2  \\ 
U_{-1} \phi_1 &= &  4 (2+\lambda) (3+\lambda)L_{-1} \phi_1  \qquad 
& U_{-1} \phi_2 &=&  4 (2-\lambda) (3-\lambda) L_{-1} \phi_2  \ . 
\end{array}
\ee
At conformal dimension two the representation has two linearly independent states, which we may take
to be $L_{-1}^2\phi_j$ and $L_{-2}\phi_j$. Expressing $W_{-2}\phi_j$ or $U_{-2}\phi_j$ 
in terms of these states and demanding
that the difference is annihilated by $L_1$ and $L_2$ fixes the corresponding coefficients. The coefficient
proportional to $L_{-2}$ turns out to be proportional to $c^{-1}$, and thus in the 't~Hooft limit we are
interested in, can be dropped.\footnote{A similar phenomenon was recently observed  in a somewhat
different context in \cite{Bertin:2011jk}.} Then we find
\be
\begin{array}{rclrcl}
W_{-2} \phi_1 &= &  - 6 L_{-1}^2 \phi_1  \qquad 
& W_{-2} \phi_2 &= & 6  L_{-1}^2 \phi_2  \\ 
U_{-2} \phi_1 &= &  10 (3+\lambda)L_{-1}^2 \phi_1  \qquad 
& U_{-2} \phi_2 &=&  10 (3-\lambda) L_{-1}^2 \phi_2  \ . 
\end{array}
\ee
By a similar argument we also find the null vectors at level three
\be
U_{-3} \phi_1 = 20 L_{-1}^3 \phi_1 \ , \qquad  \qquad
U_{-3} \phi_2 = 20 L_{-1}^3 \phi_2 \ .
\ee

\subsubsection{The highest weight calculation}

For the calculation of the highest weight space we can quotient out by the states of the form 
\begin{eqnarray}
L_{-1} (\psi_1\otimes \psi_2) & =  & 
((L_{-1} \psi_1)\otimes \psi_2) +  (\psi_1\otimes ( L_{-1}\psi_2)) \ , \\
W_{-2} (\psi_1\otimes \psi_2) & =  & 
((W_{-2} \psi_1)\otimes \psi_2) +  (\psi_1\otimes ( W_{-2}\psi_2)) \ , \\
W_{-1} (\psi_1\otimes \psi_2) & =  & 
((W_{-2} \psi_1)\otimes \psi_2) +  ((W_{-1} \psi_1)\otimes \psi_2) +  
(\psi_1\otimes ( W_{-1}\psi_2)) \ , 
\end{eqnarray}
where $\psi_1$ and $\psi_2$ are arbitrary states in ${\cal H}_1$ and ${\cal H}_2$, respectively.
These identities can be obtained from standard contour deformation arguments, see for example
\cite{Gaberdiel:1993td} for a derivation in the general case. 
Combining the last identity with the null vector relations we then conclude that in 
$({\cal H}_1\otimes {\cal H}_2)^{(0)}$ we have the relation
\begin{eqnarray}
(L_{-1}^2 \phi_1) \otimes \phi_2 & \cong & - \tfrac{1}{6} (W_{-2} \phi_1) \otimes \phi_2 \nonumber \\
& \cong & + \tfrac{1}{6} (W_{-1} \phi_1) \otimes \phi_2 + \tfrac{1}{6}  \phi_1 \otimes (W_{-1} \phi_2) \nonumber \\
& \cong & - \tfrac{1}{2} \, (2+\lambda) (L_{-1} \phi_1) \otimes \phi_2 
+ \tfrac{1}{2} (2-\lambda)  \, \phi_1 \otimes (L_{-1} \phi_2) \\
& \cong & - \tfrac{1}{2}\,  (2+\lambda) (L_{-1} \phi_1) \otimes \phi_2 
- \tfrac{1}{2} \, (2-\lambda)  (L_{-1}\phi_1) \otimes  \phi_2 = - 2 (L_{-1}\phi_1) \otimes  \phi_2 \ . \nonumber
\end{eqnarray}
This suggests that $({\cal H}_1\otimes {\cal H}_2)^{(0)}$ is spanned by the two states
\be
\bigl( {\cal H}_1 \otimes {\cal H}_2 \bigr)^{(0)}  = {\rm span} \bigl\{ e_1 = (\phi \otimes \phi) \ , \ 
e_2 = (L_{-1}\phi \otimes \phi)    \bigr\} \ .
\ee
The action of  $L_0$ is defined to be 
\be
L_{0} (\psi_1\otimes \psi_2)  = ((L_{-1} \psi_1)\otimes \psi_2) + ((L_{0} \psi_1)\otimes \psi_2)
+ (\psi_1\otimes ( L_{0}\psi_2)) \ , 
\ee
and thus we find on $( {\cal H}_1 \otimes {\cal H}_2 )^{(0)}$ 
\be
L_0 e_1  =  e_1 + e_2 \ , \qquad \qquad 
L_0 e_2  =  2 e_2 + (L_{-1}^2 \phi_1) \otimes \phi_2 \cong 0 \ .
\ee
The corresponding eigenvalues and eigenvectors are
\be
\begin{array}{ll}
h=0: \qquad \qquad & \omega^{(0)} = e_2 \\
h=1: \qquad & \psi^{(0)} = e_1 + e_2 \ .
\end{array}
\ee
Similarly, we calculate from
\be
W_{0} (\psi_1\otimes \psi_2)  = ((W_{-2} \psi_1)\otimes \psi_2) +  2 ((W_{-1} \psi_1)\otimes \psi_2)
+ ((W_{0} \psi_1)\otimes \psi_2)
+ (\psi_1\otimes ( W_{0}\psi_2)) 
\ee
the action of $W_0$ on these states to be
\begin{eqnarray}
W_0 e_1 & = & - 6 (L_{-1}^2 \phi_1) \otimes \phi_2 - 6 (2+\lambda)  (L_{-1} \phi_1) \otimes \phi_2 
- 6 \lambda e_1 \nonumber \\
& = &  - 6 \lambda e_1 + 12 e_2 - (12 + 6 \lambda) e_2 = - 6\lambda (e_1 + e_2) 
\end{eqnarray}
and
\begin{eqnarray}
W_0 e_2 \hspace*{-0.2cm} & = \hspace*{-0.2cm} 
& (L_{-1} W_{-2} \phi_1 )\otimes \phi_2 + 2 (W_{-2} \phi_1 )\otimes \phi_2 + 
2 (L_{-1} W_{-1} \phi_1 )\otimes \phi_2 + 2 (W_{-1} \phi_1 )\otimes \phi_2 - 6 \lambda e_2\nonumber \\
& = \hspace*{-0.2cm} 
& - 6 (L_{-1}^3 \phi_1 )\otimes \phi_2 + (24  - 6 (2+\lambda) -6\lambda ) e_2 
- 6 (2+\lambda) (L_{-1}^2 \phi_1 )\otimes \phi_2 \nonumber \\
& = \hspace*{-0.2cm} 
& ( -36 + 24 - 6(2+\lambda) - 6 \lambda + 12 (2+\lambda) e_2 = 0  \ ,
\end{eqnarray}
where we have used that
\be
[L_m,W_n] = (2m-n) W_{m+n} \ ,
\ee
as well as (see eq.~(\ref{A1}) in the appendix)
\be
(L_{-1}^3 \phi_1 )\otimes \phi_2 \cong - 3  (L_{-1}^2 \phi_1) \otimes \phi_2 \cong 6 e_2 \ .
\ee
Thus we conclude that 
\be\label{wvalff}
W_0 \, \omega^{(0)} = 0 \ , \qquad W_0 \psi^{(0)} = - 6 \lambda \, \psi^{(0)} \ .
\ee

\subsubsection{Going up to level one}

This calculation already implies that the state $\psi$ at $h=1$ is not any descendant of the state 
$\omega$ at $h=0$. Naively this suggests that the representation should be just the direct sum of the two
representations, but there is more to this. In fact, using similar techniques--- the details are described in the 
appendix  --- the level one space turns out to be generated by 
\be\label{4.26}
 f_1 = \phi_1 \otimes \phi_2 \ , \quad  f_2 = (L_{-1}\phi_1)\otimes \phi_2 \ , \quad
f_3 = \phi_1 \otimes (L_{-1} \phi_2) \ , \quad  f_4 = (L_{-1}^2\phi_1)\otimes \phi_2  \ . 
\ee
Note that in order to go from $( {\cal H}_1 \otimes {\cal H}_2 )^{(1)}$ to
$( {\cal H}_1 \otimes {\cal H}_2 )^{(0)}$, we have to impose
the additional relations 
\be\label{addrel}
f_1= e_1 \ , \qquad f_2 = e_2 \ , \qquad f_3 = - e_2 \ , \qquad f_4 = -2 e_2 \ . 
\ee
Using the results of the appendix we find that the eigenvectors of $L_0$ are now given by 
\be\label{4.28}
\begin{array}{ll}
h=0: \qquad \qquad & \omega^{(1)} = - \tfrac{1}{2} f_4 \\
h=1: \qquad & \psi^{(1)} = f_1 - f_2 - f_4 \\
h=2: \qquad & \rho^{(1)} = f_2 + f_3 \\
& \xi^{(1)} = 2 f_2 + f_4 \ .
\end{array}
\ee
Note that upon imposing eq.~(\ref{addrel}) $\omega^{(1)} \cong  \omega^{(0)}$ and
$\psi^{(1)}\cong \psi^{(0)}$. The additional two states, $\rho^{(1)}$ and $\xi^{(1)}$ are 
descendants, and therefore do not appear in $\bigl( {\cal H}_1 \otimes {\cal H}_2 \bigr)^{(0)}$,
{\it i.e.}\ they vanish upon imposing eq.~(\ref{addrel}). Indeed, we have 
\be\label{onedes}
L_{-1} \psi^{(1)} = f_2 + f_3 = \rho^{(1)} \qquad \hbox{and} \qquad
W_{-1}\psi^{(1)} = -6 \xi^{(1)} + (6-3\lambda) \rho^{(1)} \ .
\ee
One also observes that the above answer is in agreement with what one expects based on the 
character (\ref{deco}): the states $\omega$ and $\psi$ are the leading states in $\chi_{(0;0)}$, 
as well as in the second sum, respectively. The vacuum character does not have any descendants 
at level one, whereas the second term has an expansion of the form $q(1 + 2 q + \cdots)$, thus 
leading to two descendants at level one which we can identify with $\rho$ and $\xi$, in agreement
with eq.~(\ref{onedes}).

Now we come to the central part of the calculation. Since we have determined the level one space
$( {\cal H}_1 \otimes {\cal H}_2 )^{(1)}$ we can work out $L_1$ on $\psi$, and check whether it
vanishes --- this would be the case if the fusion product is the direct sum of the two representation. 
Using that 
\be\label{4.30}
L_1 (\psi_1\otimes \psi_2) = ((L_{-1} \psi_1 ) \otimes \psi_2) 
+ 2  ((L_{0} \psi_1 ) \otimes \psi_2) +  ((L_{1} \psi_1 ) \otimes \psi_2) +  (\psi_1 \otimes (L_{1} \psi_2))
\ee
we find 
\be\label{4.31}
\begin{array}{rclrcl}
L_1 \omega^{(1)} & = & 0 \qquad \qquad & L_1 \psi^{(1)} & = & - \lambda\, \omega^{(0)} \\
L_1 \rho^{(1)} & = & 2 \psi^{(0)} \qquad \qquad & L_1 \xi^{(1)} & = & 2 (1+\lambda) \psi^{(0)} \ .
\end{array}
\ee
The crucial identity is the second relation in the first line: it shows that, for $\lambda\neq 0$,  
$\psi$ is not annihilated by $L_1$, but rather is mapped to $\omega$.  The resulting representation
has therefore the schematic structure of (\ref{ffdia}). Thus while $\omega$ can be obtained by the action
of a mode from $\psi$, we cannot obtain $\psi$ from $\omega$ by the action of any (negative) mode. 
The resulting representation is therefore reducible --- the states that are generated by the action of 
the (negative) modes from $\omega$ form a proper subrepresentation --- but we cannot decompose 
the representation completely, {\it i.e.}\
it cannot be written as a direct sum of irreducible representations. In fact, we cannot write the 
representation as a direct sum of any two smaller representations. Reducible representations with
this property are often referred to as {\em indecomposable} representations.

We have also done the analogous analysis for the Virasoro case, {\it i.e.}\ for the fusion of the 
two $h=\tfrac{1}{4}$ representations (that correspond to $(2,1)$ and $(1,2)$ in the $c\rightarrow 1$ 
limit), and in that case the representation actually decomposes. This fits in nicely with the above analysis
since the $c\rightarrow 1$ limit of the Virasoro minimal models corresponds  formally to $\lambda=0$,
in which case also the above representation becomes just the direct sum of two 
irreducible representations.

\subsection{Other Cases}\label{ss:othercases}

One may be worried that the phenomenon we have just described is only a consequence of the fact
alluded to before, that we cannot obtain the vacuum in the fusion of two states with different conformal 
dimensions. We have therefore checked other cases, where this argument does not apply. In particular,
we have studied the fusion of $({\rm f};0)\otimes (0;\tableau{1 1})$, where the product has states of
conformal dimensions $h=\tfrac{1}{2}(1-\lambda) + n$ with $n\in{\mathbb N}$, {\it i.e.}\ none of the
states in question has $h=0$. The structure of the resulting representation is 
\be\label{ffdiab}
(\ff; \tableau{1 1})\!: \qquad 
\xymatrix@C=1pc@R=2.2pc{
  & \vdots & & \vdots & & \vdots & & \vdots \\
 & \half(5-\lambda) & & \ar@<0.4ex>[dr]^{L_1}\pi & & \sigma\ar@<-0.4ex>[dl] &  & \\
 & \half(3-\lambda) & & &\ar[dr]_{L_1}\ar@<0.4ex>[ul] \nu\ar@<-0.4ex>[ur] & & \ar@<-0.4ex>[dl]_{L_1}\rho & \\
 L_0= & \half(1-\lambda) & & & &\ar@<-0.4ex>[ur]_{L_{-1}}\mu& & 
 }
\ee 
Again, this is an indecomposable representation since the states that are obtained from the highest weight
state $\mu$ by the action of the (negative) modes only form a subrepresentation since one cannot obtain 
$\nu$ as a descendant of $\mu$. The unlabeled arrow from $\nu$ to $\sigma$ is a linear combination of 
$L_{-1}$ and $W_{-1}$. The details of the calculation are described in appendix~\ref{sec:4.2.1}.

Note that also in this case the decoupling argument of section~\ref{ss:decouplingb} goes through.
While there is now a descendant of $\mu$ (namely $\rho$) that could potentially have a non-zero
two-point function with $\nu$, the two-point function is actually zero since $\nu$ and $\rho$ have
different $W_0$ eigenvalues, see eq.~(\ref{W0ei}).
\medskip

In order to double check our analysis, we have also studied an example where we {\em do} 
expect the resulting representation to be completely decomposable, namely the fusion 
\be
({\rm f};0) \otimes ({\rm f};0)  = (  \tableau{2} ;0) \oplus    (\tableau{1 1};0)  \ . 
\ee
The calculation is described in appendix~\ref{ss:doubleblind}, and the resulting representation is 
indeed the direct sum.
\medskip

\noindent Finally, we have studied the example 
\be
(\tableau{1 1}; \tableau{1 1}) = 
(\tableau{1 1};0) \otimes (0;\tableau{1 1}) \ ,
\ee
where one naively expects a decomposition into three representations. The details of 
the calculation are described in appendix~\ref{sec:4.2.3}; the resulting structure is precisely
as expected, 
\be\label{ffdiac}
(\tableau{1 1}; \tableau{1 1})\!: \qquad
\xymatrix@C=.8pc@R=1.5pc{
& \vdots & & & \vdots & & \vdots & & & \\
& 3 & & & \pi\ar@<-.4ex>[dr] & & \chi\ar@<.4ex>[dl] & & \vdots &  \\
& 2 & & & & \rho \ar[drr]_{L_1} \ar@<-.4ex>[ul] \ar@<.4ex>[ur]& & \sigma\ar@<-.4ex>[d] & & \xi\ar@<.4ex>[dll]  \\
& 1 & & & & & & \psi\ar@<-.4ex>[u] \ar[d]^{L_1}\ar@<.4ex>[urr]& &  \\
L_0= & 0 & & & & & &\omega& &  
 }
\ee 
Again the decoupling argument of section~\ref{ss:decouplingb} goes through since $\rho$
cannot have a non-zero two-point function with either $\sigma$ or $\xi$ since their $W_0$ 
eigenvalues are again different, see eq.~(\ref{W001}). 

\section{The Role of the Higher Spin Algebra}\label{s:hsalgebra}

In the previous sections we demonstrated that $Z_{\rm{bulk}} = Z_{\rm{CFT}}$ by rewriting both partition 
functions as sums over ${\mathfrak u}(\infty)$ characters.  This is somewhat mysterious since 
neither side has any obvious ${\rm U}(N)$ symmetry.  To explain why these characters arise, 
we will show that they can be reinterpreted as characters of the higher spin algebra \hs{\lambda} that 
governs the bulk gauge theory. Thus it is obvious that some reorganization of the bulk partition function 
into ${\mathfrak u}(\infty)$ characters should be possible, as done in section \ref{s:gravity}.  Then, the 
rest of this section is devoted to providing evidence that the $\W_N$ symmetry of the CFT,  in the 't~Hooft limit, is equivalent to the \hs{\lambda} symmetry of the bulk. 
If we assume this equivalence then the fact that both partition functions can be written in terms of  \hs{\lambda} characters gives 
a {\it partial} explanation of their agreement since it constrains both partition functions considerably. However, these considerations  by themselves are obviously not sufficient for a proof since 
they do not imply that the same representations appear with the same multiplicity.

\subsection{Characters of \hs{\lambda}}

The higher spin algebra \cite{Feigin88,Bordemann:1989zi,Bergshoeff:1989ns,Vasiliev:1989re,Pope:1989sr} (see \cite{Gaberdiel:2011wb} for a review and further references) can be obtained from the 
universal enveloping algebra of $\mathfrak{sl}(2)$.  Specifically, denoting the  $\mathfrak{sl}(2)$ generators 
by $J_0$, $J_\pm$, the generators of \hs{\lambda} are polynomials in 
$J_{0,\pm}$ modulo the relation that the Casimir equals
\be\label{valcas}
C_2 \equiv J_0^2 - \half (J_+ J_- + J_- J_+) = \frac{1}{4}(\lambda^2 - 1) \ .
\ee
A basis for the resulting quotient space is given by the higher spin generators 
\be\label{env}
V^s_m = n_s\, (-1)^{m+1}(m+s-1)! \, 
 \Bigl[ \underbrace{J_- , \dots [J_-, [J_-}_{\hbox{\footnotesize{$s-1-m$ terms}}}, J_+^{s-1}]]\Bigr] \ , \quad
 s\geq 2 \ , \quad |m|<s \ ,
\ee
with $n_s$ a normalization factor; to avoid constants in the eventual comparison to $\W_N$, we choose 
$n_2 = \half$ and for $s>2$,  $n_s =((s-1)!)^{-2}$. Then the spin-2 generators 
$L_{m}\equiv V^{s=2}_{m}$  for $m=0,\pm 1$ simply agree with $J_{0,\pm}$. 

The simplest representations of \hs{\lambda} are those inherited from representations of 
$\mathfrak{sl}(2)$ with Casimir equal to (\ref{valcas}). In particular, we may consider
the highest-weight representation of $\mathfrak{sl}(2)$ with highest weight $J_0=h$, for which
the condition on the Casimir translates into $h=h_{\pm}=\tfrac{1}{2}(1\pm \lambda)$, see eq.~(\ref{hpm}). 
Since $L_0=V^{s=2}_{0} = J_0$, its trace equals 
\be\label{hsfund}
\tr_{\ff} \, q^{L_0} = \frac{q^h}{1-q} \ .
\ee
There are actually four such representations since for each choice of $h=h_\pm$ there is a pair
of  conjugate representations that differ by the sign of the eigenvalues of the odd spin generators. 
Let us denote these representations by $\ff_\pm$ and $\bar{\ff}_\pm$, respectively.  Note that
these characters agree precisely with those of the corresponding ${\mathfrak u}(N)$
representation
\be
\Tr_{\ff_\pm} q^{L_0}  = \Tr_{\bar{\ff}_\pm} q^{L_0}   = P_{\ff}^\pm(q)\ .
\ee

\subsubsection{Other representations of \hs{\lambda}}

Obviously, the above four representations are not the only representations of \hs{\lambda}. Indeed,
we can generate additional representations by taking tensor products. Many of these
tensor products are actually irreducible by themselves. For example, let $\H_1$
and $\H_2$ be two such representations, and consider $\phi_1\in \H_1$ and $\phi_2\in\H_2$. 
Then we have
\bea
V^2_{-1} (\phi_1 \otimes \phi_2) & = & ( J_{-} \phi_1 ) \otimes \phi_2 + \phi_1 \otimes (J_{-} \phi_2) \\
V^3_{-1} (\phi_1 \otimes \phi_2) & = &  \mp 3(1+2h_1) \, 
( J_{-} \phi_1 ) \otimes \phi_2) \mp  3(1 + 2h_2)\, \phi_1 \otimes (J_{-} \phi_2)  \ ,
\eea
where the signs in the last line correspond to whether $\H_i$ is the representation
corresponding to ${\rm f}_{\pm}$ or $\bar{\rm f}_{\pm}$, and $h_i$ is the $J_0$ eigenvalue of $\phi_i$. 
Unless $\H_1=\H_2$, it is then clear that a suitable linear combination of $V^2_{-1} (\phi_1 \otimes \phi_2)$ 
and $V^3_{-1} (\phi_1 \otimes \phi_2)$
will be equal to $( J_{-} \phi_1 ) \otimes \phi_2$, while another one will be equal to 
$\phi_1 \otimes (J_{-} \phi_2)$. Thus it follows that the representation $\H_1\otimes \H_2$
is irreducible. 

This argument obviously breaks down for $\H_1=\H_2$. In this case, the
tensor product is not irreducible. In fact, it is clear that the permutation group commutes
with the Lie algebra action on the tensor product, and hence we can decompose the tensor 
product by the usual Young tableau technology. For example, for the two-fold tensor product we 
simply get
\be\label{twode}
\H \otimes \H = \H_{\tableau{2}} \oplus \H_{\tableau{1 1}} \ ,
\ee
where $\H$ is any of the four representations ${\rm f}_{\pm}$ and $\bar{\rm f}_{\pm}$,
and we parametrize the representations of \hs{\lambda} by the familiar Young tableaux.
(Obviously there are four versions of these representations, depending on whether $\H$ is
${\rm f}_{\pm}$ or $\bar{\rm f}_{\pm}$.) In any case, the characters of the two irreducible 
representations  that appear in (\ref{twode}) are 
\be\label{twoch}
\tr_{\tableau{2}} q^{L_0} = \frac{q^{2h}}{(1-q)(1-q^2)} \ , \qquad
\tr_{\tableau{1 1}} q^{L_0} = \frac{q^{2h+1}}{(1-q)(1-q^2)}  \ ,
\ee
where $h=h_\pm$. Note that these characters satisfy 
\be\label{6.9}
\bigl(\tr_{\ff} \,q^{L_0}\bigr) ^2 = \frac{q^{2h}}{(1-q)^2} = 
\frac{q^{2h}}{(1-q)(1-q^2)} +  \frac{q^{2h+1}}{(1-q)(1-q^2)}  = 
 \tr_{\tableau{2}} q^{L_0} + \tr_{\tableau{1 1}} q^{L_0} \ .
\ee
Because they count the symmetrized and anti-symmetrized contributions, they agree again
with the corresponding ${\mathfrak u}(N)$ characters. The same argument obviously generalizes
to arbitrary tensor powers, and we therefore conclude that 
\be
\Tr_{R_\pm} q^{L_0} = P_{R_\pm}^\pm(q) \ ,
\ee
where $R_{\pm}$ corresponds to a Young tableaux with finitely many boxes. 
A general representation of \hs{\lambda} obtained from the four fundamentals is then labeled by 
four Young tableaux, labeled according to 
\be
\ff_\pm \leftrightarrow R_\pm \ , \qquad \bar{\ff}_\pm \leftrightarrow S_{\pm}  \ ,
\ee
and its character equals 
\be
\mbox{Tr}_{R_+,R_-,S_+,S_-} q^{L_0} = P_{R_+}^+(q) P_{S_+}^+(q) P_{R_-}^-(q) P_{S_-}^-(q)\ .
\ee
Note that this is precisely what appears in eq.~(\ref{gravityfinal1}).

\subsection{\hs{\lambda} and $\W_N$}

Next we want to explain why the CFT partition function in the 't~Hooft limit can also be written in terms of 
\hs{\lambda}  representations. On the face of it, this is much more mysterious than the analysis in the bulk,
as we shall now explain. On general principles, the symmetry algebra of the boundary CFT can be 
identified with the asymptotic symmetry algebra of the gravity (or higher spin) theory, and it will usually 
form some $\W$ algebra. The `global' symmetry of the CFT, {\it i.e.}\ the analogue of \hs{\lambda}, can 
then be identified with the wedge algebra,  {\it i.e.}\ the algebra generated by the modes 
$W^s_n$ with $|n|<s$, where $s$ denotes the  spin of $W^s$. This algebra 
forms a proper subalgebra of the full $\W$ symmetry at $c\rightarrow \infty$ \cite{Bowcock:1991zk}. 

For the case at hand, the local symmetry of the bulk higher spin theory is, of course, two 
copies of \hs{\lambda}.  The bulk asymptotic  symmetries form (two copies of) a larger algebra 
\w{\lambda} that extends \hs{\lambda} much like the  Virasoro algebra extends $\mathfrak{sl}(2)$
\cite{Henneaux:2010xg,Campoleoni:2010zq,Gaberdiel:2011wb}.  \w{\lambda} has an infinite 
number of conserved currents $W^s$ of spins $s=2,3,\dots$, where $W^2$ is the stress tensor. 
This is the higher spin analogue of the construction of the Virasoro algebra in ordinary 
3d gravity by Brown and Henneaux \cite{Brown:1986nw}. The asymptotic symmetry computation is, 
algebraically, identical to the classical Drinfeld-Sokolov (or Hamiltonian) reduction.  DS reduction 
is a purely algebraic procedure to construct a $\W$-algebra starting with any Lie algebra.  For the 
case at hand, the Lie algebra in question is \hs{\lambda}, and the DS point of view guarantees that
the wedge subalgebra of \w{\lambda} agrees again with \hs{\lambda}. This is then in perfect
agreement with what we expect from the bulk point of view.
\smallskip

However, in the proposal of \cite{Gaberdiel:2010pz}, the dual CFT was {\em not} described in terms
of \w{\lambda}, but rather as the 't~Hooft limit of the $\W_N$ minimal models. At finite $N$, the wedge 
subalgebra of $\W_N$ equals $\mathfrak{sl}(N)$, and thus one may naively think that the wedge 
subalgebra  of the 't~Hooft limit is something like $\mathfrak{sl}(\infty)$, which on the face of it
would be different from \hs{\lambda}. In the following we want to argue that the wedge subalgebra of
the 't~Hooft limit is equal to \hs{\lambda}. (Actually, our arguments also suggest that 
the full 't~Hooft limit of the $\W_N$ minimal models has  \w{\lambda} symmetry.)
In particular, this then implies that the states of the 't~Hooft limit  theory form representations of 
\hs{\lambda}, mirroring what we have seen above for the bulk.

Unfortunately, it is not possible to determine the structure of the wedge subalgebra directly, since the 
commutation relations of $\W_N$ are known only at small $N$, and the details of the large $N$ limit 
are sensitive to a nonlinear choice of basis.  Some evidence based on the representation theory of the 
two algebras was already given in \cite{Gaberdiel:2011wb}; here we will not give a complete proof, but 
motivate the conjecture from level-rank duality and check it explicitly by comparing properties of many 
degenerate representations of the two algebras, thereby extending the checks of \cite{Gaberdiel:2011wb}
considerably.

\subsubsection{Level-rank duality}

The proposed duality between the 't~Hooft limit of the $\W_N$ minimal models and \w{\lambda} is the
natural generalization of a certain class of level-rank dualities for coset models that were discovered
some time ago by \cite{Kuniba:1990zh,Altschuler:1990th}. It was
observed in these papers that for coprime integers $(M,N)$, the  two cosets 
\be
{\mathfrak{su}(N)_k \oplus \mathfrak{su}(N)_1 \over \mathfrak{su}(N)_{k+1}} \ 
\cong \  
{\mathfrak{su}(M)_l \oplus \mathfrak{su}(M)_1\over \mathfrak{su}(M)_{l+1}} \
\ee
are equivalent, where the levels are fractional  and defined by 
\be
M = {N\over k+N} \ , \qquad l = {M\over N} - M \ .
\ee
If we take $N$ large and blindly analytically continue to non-integer rank $M \rightarrow \lambda \in [0,1]$, 
then the left-hand side is the $\W_N$ minimal model at $c = N(1-\lambda^2)$, corresponding to the 't Hooft 
limit.  The symmetry associated to the right-hand side is the Drinfeld-Sokolov reduction of 
$\mathfrak{sl}(\lambda)$ at the same value of the central charge.  But the higher spin algebra is known to 
arise as an analytic continuation of $\mathfrak{sl}(N)$ to non-integer rank $N \rightarrow \pm \lambda$ 
\cite{Feigin88,Fradkin:1990qk}; this suggests that the proper definition of the analytic continuation on 
the right-hand side gives precisely \w{\lambda}, the DS reduction of \hs{\lambda}. 

This intuition was made more precise and used to relate various quantum $\W$-algebras in 
\cite{Blumenhagen:1994ik,Blumenhagen:1994wg,Hornfeck:1994is}; our conjecture 
is a classical limit thereof.  It would be interesting to understand what the quantum results of  
\cite{Blumenhagen:1994ik,Blumenhagen:1994wg,Hornfeck:1994is} imply for the duality at finite 
$N$, where the bulk symmetries should presumably be obtained by a quantum Drinfeld-Sokolov 
reduction.

\subsubsection{Comparison of degenerate representations}

As further evidence of the equivalence of the 't~Hooft limit of $\W_N$ with \w{\lambda} 
we now compare the eigenvalues of some $W^s_0$ generators on primary states. Let us 
first recall the description of primary states for the two cases. For the $\W_N$ minimal model, 
the interesting ({\it i.e.}\ degenerate) representations are labeled by 
\be
\Lambda = \alpha_+\Lambda_+ + \alpha_-\Lambda_- \ ,
\ee
where
\be
\alpha_+\alpha_- = -1 \  , \qquad  \alpha_- = - \sqrt{k_{\rm DS}+N}\ , \qquad 
\alpha_0 = \alpha_+ + \alpha_- \  ,
\ee
with $k_{\rm DS}$ the level of the DS-reduction, 
which is related to the level $k$ in the coset description via
\be\label{krel}
\frac{1}{k+N} = \frac{1}{k_{\rm DS}+N} - 1  \ .
\ee
Furthermore, $\Lambda_+$ and $\Lambda_-$ are representations of $\mathfrak{su}(N)$ with
Dynkin labels $\Lambda_\pm = \sum_s \Lambda_s^\pm\, \lambda_s$, 
where $\lambda_s$, $s=1,\dots, N-1$ are the fundamental weights.  Equivalently, we can describe 
$\Lambda_\pm$ in terms of the associated Young tableaux that are most conveniently characterized by
the number of boxes $r_j^\pm$ in the $j^{\rm th}$ row. The eigenvalues of the primary $W^s_0$ generators
\be
W^s_0 |\Lambda\rangle = w^{(s)}(\Lambda) |\Lambda\rangle
\ee
are explicitly known for $s=2,3,4$. (An explicit formula for the eigenvalues of the non-primary generators
$U^s_0$ is known in closed form for all $s$; however, the transformation to the primary basis of $W^s_0$
is only available for the first few $s$ --- see appendix~\ref{a:hsdetails} for more details.) For example, 
the conformal dimension $h(\Lambda) \equiv w^{(2)}(\Lambda)$ equals 
\be\label{hLam}
h(\Lambda) = \frac{1}{2} (\Lambda, \Lambda + 2 \alpha_0 \rho) \ .
\ee
Explicit descriptions for $w^{(s)}(\Lambda)$, $s=3,4$, are given in appendix~\ref{a:hsdetails}.
\smallskip

Next we want to deduce from this general formula the eigenvalues in the two different cases that are 
of interest to us. In the 't~Hooft limit, we take $N,k\rightarrow \infty$, keeping $\lambda=\tfrac{N}{k+N}$ fixed. 
Rewriting (\ref{krel}) as 
\be
\frac{N}{N+k} = \lambda = \frac{N}{k_{\rm DS}+N} - N
\ee
this means that we take $N\rightarrow \infty$, with $k_{\rm DS} = - N +1 - \tfrac{\lambda}{N}$.  In terms of
$\alpha_\pm$ and $\alpha_0$ this becomes 
\be\label{tHooftalpha}
\hbox{'t~Hooft limit:} \quad 
\alpha_+ \cong +1 \ , \qquad \alpha_- \cong -1 \ , \qquad \alpha_0 \cong \tfrac{\lambda}{N} \ .
\ee
For example, for the conformal dimension $h(\Lambda)$, we then obtain in this limit
\be\label{htH}
\hbox{'t~Hooft limit:} \quad  h_{\rm mm}(\Lambda) = \frac{1}{2} \sum_{j} r_j^2 +\frac{1}{2} \lambda  B \ , 
\ee
where we have used the relations of appendix~\ref{a:orthobasis}, see in particular
eqs.~(\ref{A.8}) and (\ref{A.11}).
\smallskip

In order to obtain the eigenvalues for \hs{\lambda} we proceed somewhat indirectly, using
the fact \cite{Feigin88,Fradkin:1990qk} that \hs{\lambda} is the analytic continuation of 
$\mathfrak{sl}(N)$ to non-integer $N\rightarrow \pm \lambda$. Thus we can compute the eigenvalues using
the above DS-reduction, but evaluated at $N = - \lambda$ and $k_{\rm DS}\rightarrow \infty$ 
\be\label{hsalpha}
\hbox{\hs{\lambda}}: \qquad
\alpha_+ \cong \frac{1}{\sqrt{k_{\rm DS}}} \ ,  \qquad 
\alpha_- \cong - \sqrt{k_{\rm DS}} \ , \qquad \alpha_0 \cong - \sqrt{k_{\rm DS}} \ .
\ee
Evaluating (\ref{hLam}) in this limit, we note that the limit only converges provided that
$\Lambda_-=0$.\footnote{We have checked that the agreement between the eigenvalues
also works for representations of the form $(0;\Lambda_-)$. In fact, that case reduces to the
above by changing the sign of $\lambda$.} Writing $\Lambda=\Lambda_+$, we then find 
\be
h_{\rm hs}(\Lambda) =  -  (\Lambda,\rho) = 
\frac{1}{2} \sum_j c_j^2 - \frac{1}{2} BN \ ,
\ee
where we have used (\ref{A.8}) and (\ref{A.11}). Replacing $N\mapsto -\lambda$ this becomes
\be\label{hhs}
\hbox{\hs{\lambda}}: \qquad h_{\rm hs}(\Lambda) = \frac{1}{2} \sum_j c_j^2 + \frac{\lambda B}{2} \ .
\ee
Comparing (\ref{htH}) with (\ref{hhs}) we now conclude that 
\be
h_{\rm mm}(\Lambda) = h_{\rm hs}(\Lambda^T) \ , 
\ee
where $\Lambda^T$ is the representation obtained from $\Lambda$ upon transposing the
Young tableaux. The appearance of the transpose in this relation is very natural, since we can
think of the correspondence as a sort of level-rank duality. 

We have similarly checked the relation for $s=3,4$, and in all cases we have for $\Lambda=(\Lambda;0)$
\be\label{wsid}
w^{(s)}_{\rm mm}(\Lambda) = w^{(s)}_{\rm hs}(\Lambda^T) \ .
\ee
The details of these calculations are presented in appendix~\ref{a:hsdetails}.  
Taken together this gives very strong support to the claim that the 't~Hooft limit of the $\W_N$ 
minimal models defines a \w{\lambda} theory, as suggested by the bulk analysis.  

Here we have checked the eigenvalues by analytic continuation from $\mathfrak{sl}(N)$, but there is 
another way to do the computation directly in \hs{\lambda}, at least for simple representations.  The
wedge modes, {\it i.e.}\ the modes $W^s_m$ with $|m|<s$, of the $\W$-algebra that is 
obtained by DS reduction from \hs{\lambda} define for $c\rightarrow \infty$ a subalgebra that is identical
to \hs{\lambda} itself.  Therefore zero modes of the $\W$-algebra can be computed using the description 
of \hs{\lambda} as the quotient of $U(\mathfrak{sl}(2))$. We have checked that for some simple 
representations, this reproduces the answers above.  The advantage of this method is that it can 
be applied to representations of the type $(\Lambda_+; \Lambda_-)$, whereas the analytic continuation 
is straightforward only for $(\Lambda_+; 0)$ or $(0; \Lambda_-)$ with $\Lambda_\pm$ a finite tensor 
power of fundamentals or anti-fundamentals.  Consider, for example, $(\ff; \ff)$.  In \hs{\lambda}, 
it corresponds to the representation formed by the tensor product $(\ff_+\otimes \bar\ff_-)$, for which
the zero mode eigenvalues add,
\be
V^s_0 \bigl( \ff_+ \otimes \bar\ff_-  \bigr)= (V^s_0 \ff_+) \otimes \bar\ff_- + \ff_+ \otimes (V^s_0 \bar\ff_-) \ .
\ee
Thus the conformal weight and spin-3 eigenvalue are
\be
h = 1 \ , \quad w^{(0)} = -6\lambda \ .
\ee
This agrees with (\ref{wvalff}) for the level-1 state $\psi$ inside the $(\ff; \ff)$ representation of $\W_N$ 
in the 't~Hooft limit, \textit{i.e.}  the ground state of the irreducible representation formed after 
removing the null vector at level 0.  We have also checked this for the other explicit examples we
have considered, see appendix~\ref{a:fusiondetails}, and we expect this pattern to continue for
all other representations $(\Lambda_+; \Lambda_-)$: eigenvalues 
derived from the higher spin algebra agree precisely with $\W_N$ as $N\rightarrow \infty$ 
after the null states that are described in section \ref{s:fusion}  are  modded out.


\section{Discussion}\label{s:discussion}

In this paper we have shown that the perturbative spectrum of the bulk higher spin
gravity theory is in one-to-one correspondence with the states of the 
dual CFT whose Young tableaux $(\Lambda_+;\Lambda_-)$ only contain 
finitely many boxes and antiboxes. The precise mapping only works at infinite $N$,
and it relies on a novel decoupling phenomenon: certain primary states of the 
CFT become null-descendants in the limit and decouple from the spectrum. 
It is an important open problem to understand the coupling to these states at large but 
finite $N$, which could potentially modify the connected correlators of the CFT even at 
leading order in $1/N$. The answer should also shed some light on the significance of the 
condition $\lambda\neq 0$ that was important for the decoupling. 

As mentioned in the introduction, in the large $N$ limit, the CFT also contains another interesting 
class of representations 
$(\Lambda_+;\Lambda_-)$, namely those for which the number of boxes or antiboxes
grows with $N$. Some of these states have actually low-lying conformal dimension --- in particular, this is 
the case for the primaries with $\Lambda_+=\Lambda_-$ for which 
$h_{(\Lambda;\Lambda)} = \tfrac{C_2(\Lambda)}{p(p+1)}$ --- and their density at any energy 
grows exponentially with $N$. In the large $N$ limit, these representations, however, 
do not couple to the  perturbative excitations in the bulk since the fusion rules of the CFT close on 
states with finitely many Young tableaux boxes. 

However, at finite $N$, the distinction between the 
two classes of states --- those that have ${\cal O}(1)$ number of boxes and antiboxes, and 
those for which this number is of order ${\cal O}(N)$ --- becomes 
ambiguous and we should be careful about the statements on decoupling. 
The fusion rules of the CFT, which are just the tensor product rules for large enough $N$, guarantee
that $m$-fold products of the four fundamental fields do not couple to states with more than $m$ boxes
or antiboxes, and thus there is still a natural separation between the two classes of states.
It is therefore intriguing as to what these states are from a bulk point of view. One possibility is that they 
correspond to a class of exotic black hole like objects in the Vasiliev theory whose masses are 
${\cal O}(1)$ even in the large $N$ limit. A first attempt at constructing 
new black holes in higher spin theories has recently been made in  \cite{Gutperle:2011kf},
but the black holes described there do not seem to have this property. 
It has also been noted by \cite{Castro:2010ce} that in the finite $N$ theories black holes would 
have other unusual features.  The presence of a primary with maximal dimension would imply 
that very large black holes are largely comprised of generalised boundary graviton excitations. 

In Witten's recent approach to 3d pure gravity \cite{Witten:2007kt}, the bulk theory is topological, 
and the spectrum is empty (other than descendants) below the black hole threshold $c/24$.  The present 
situation is rather different: we have focused on states below $c/24$, which are nontrivial because the 
bulk has propagating degrees of freedom.  Since the higher spin theory also contains standard BTZ 
black holes, the usual Farey tail picture \cite{Dijkgraaf:2000fq,Maloney:2007ud,Manschot:2007ha} 
suggests that CFT states above the black hole threshold might be obtained from the perturbative part 
of the spectrum upon applying modular transformations. In the  $N\rightarrow \infty$ limit, the central 
charge of the CFT also goes to infinity, and modular transformations are delicate. However, it would be 
interesting to see whether one can make sense of this at finite $N$. An understanding of this issue would 
also be important in order to get insight into the thermodynamic properties of the theory, in particular, the 
question of whether a Hawking-Page transition takes place. 

Another route to understanding the higher spin theory at finite $N$ may be to exploit the connection to 
topological Chern-Simons theory.  The $\W_N$ CFT, being a coset model, is related to a 3d topological 
Chern-Simons gauge theory by the Chern-Simons/Wess-Zumino-Witten map 
\cite{Witten:1988hf} as applied to cosets \cite{Moore:1989yh}.  This 3d theory is different from higher spin gravity, which has a 
non-compact symmetry group and is not topological.  Nonetheless the higher spin characters 
$P_R^\pm(q)$ that appear in the bulk and boundary partition functions are related to knot polynomials, 
as pointed out in section \ref{s:cft}.

Finally, these investigations of the finite $N$ theory are presumably tied up with the basic question as to 
whether it is possible to have a consistent quantum mechanical version of the Vasiliev theory by itself or 
if it is necessary to embed it in a larger theory (with more degrees of freedom) such as string theory.


\bigskip

\noindent
{\bf Acknowledgements:} We are grateful to Chi-Ming Chang, Miranda Cheng, Shiraz Minwalla, Kyriakos Papadodimas, Mukund Rangamani, Misha Vasiliev, 
Carl Vollenweider, and Xi Yin for useful discussions. The work of MRG is supported 
by the Swiss National Science Foundation, and he thanks the Weizmann Institute (Rehovot) and
the Simons Center (Stony Brook) for hospitality while part of this work was being done. The work of R.G. is partially funded by the SwarnaJayanthi Fellowship of the DST, Govt. of India and more broadly by the generous support
for basic sciences by the people of India.  He would also like to thank IAS (Princeton), Simons Center (Stony Brook), IFT (Utrecht), DESY (Hamburg) and ICTS (Bangalore) for hospitality during the completion of this work. 
TH is supported by U.S. Department of Energy grant DE-FG02-90ER40542. S.R. is supported by a Ramanujan fellowship of the Department of Science and Technology.  S.R. would also like to acknowledge the support of the Harvard University Physics department and is grateful to the Perimeter Institute (Waterloo), CERN (Geneva), and ETH (Zurich) for their hospitality while this work was being completed.

\section*{Appendices} 
\appendix

\section{Orthogonal Basis for Characters}\label{a:orthobasis}

For some of these calculations it is useful to write the weights and roots in an orthogonal basis.
Let $e_i$, $i=1,\ldots, N$ be an orthonormal basis. The simple roots of $\mathfrak{su}(N)$ are then given
by $\alpha_i=e_i-e_{i+1}$, while the fundamental weights are 
\be
\lambda_i = \sum_{j=1}^{i} e_j  - \frac{i}{N} \sum_{j=1}^{N} e_j \ , \qquad i=1,\ldots, N-1 \ , 
\ee
and thus the Weyl vector is
\be
\rho = \sum_{i=1}^{N} \bigl( \tfrac{N+1}{2} - i \bigr) \, e_i \ .
\ee
The finite Weyl group is then just the permutation group $S_N$ acting on the $e_i$. 
The Dynkin labels of a representation with highest weight $\Lambda$ are the coefficients 
$\Lambda_s$  in 
\be
\Lambda = \sum_{s=1}^{N-1} \Lambda_s \, \lambda_s \ . 
\ee
Given $\Lambda_s$, the corresponding Young tableau has $r_j$ boxes in the $j^{\rm th}$ row
with
\be
r_j = \sum_{s=j}^{N-1} \Lambda_s \ , \qquad \Lambda_s = r_s - r_{s+1} \qquad
B = \sum_{j=1}^{N} r_j =  \sum_{s=1}^{N-1} s \Lambda_s \ ,
\ee
where $B$ is the total number of boxes in the Young tableau. 
It is sometimes also useful to write $\Lambda$ directly in terms of the orthonormal
basis $e_i$ itself, {\it i.e.}
\be
\Lambda = \sum_{i=1}^{N} l_i\, e_i \ .
\ee
Then, with $r_N \equiv 0$, we have 
\be
\label{orthinrows}
l_j = \sum_{s=j}^{N-1} \Lambda_s - \frac{B}{N} = r_j - \frac{B}{N} \ , \qquad \sum_{j=1}^{N} l_j = 0 \ ,
\ee
{\it i.e.}\ the orthogonal labels differ from the labels describing the length of the rows by an overall
constant that guarantees that their sum vanishes.

In terms of the orthogonal labels, the quadratic Casimir can now be easily calculated, 
\begin{eqnarray}
C_2 =\frac{1}{2} \langle \Lambda, \Lambda + 2 \rho \rangle 
& = &  \frac{1}{2} \sum_{i=1}^{N} l_i^2  + \sum_{i=1}^{N} l_i \bigl( \tfrac{N+1}{2} - i \bigr) \nonumber \\
& = &  \frac{1}{2} \sum_{i=1}^{N} r_i^2 - \frac{B^2}{2N} - \sum_{i=1}^{N} i l_i \nonumber \\
& = & \frac{1}{2} \sum_{i=1}^{N} r_i^2 - \frac{B^2}{2N} - \sum_{i=1}^{N} i r_i + \frac{1}{2} B (N+1) \ , \label{A.8}
\end{eqnarray}
where we have used that the sum of the $l_j$ equals zero. To evaluate this further, 
we denote by $c_j$ the number of boxes in the $j^{\rm th}$ column; the variables 
$c_j$ and $r_j$ are related through
\begin{equation}
c_j = \sum_{i=1}^{N-1} H(r_i - j)\ , \qquad r_j = \sum_i H(c_i - j)\ ,
\end{equation}
where $H$ is the Heaviside step function with the convention that $H(0) = 1$. In the second 
sum, $i$ is unrestricted but $r_{j \geq N} = 0$, since no column can have more than 
$N-1$ boxes. This leads to 
\begin{equation}
\sum_{j=1}^{N-1} j r_j = \sum_i \sum_{j=1}^{N-1} j H(c_i - j) 
= \sum_i {c_i (c_i + 1) \over 2} = \half\sum_i c_i^2 + {B \over 2}\ . \label{A.11}
\end{equation}
Plugging into (\ref{A.8}) it then follows that the quadratic Casimir equals
\be\label{A.10}
C_2 = \frac{1}{2} B N + \frac{1}{2} \bigl( \sum_{i=1}^{N-1} r_i^2 - \sum_{j} c_j^2 \bigr) - \frac{B^2}{2N} \ ,
\ee
thus leading to (\ref{casexp}). The Weyl denominator formula implies
\begin{eqnarray}
\sum_{w \in W} \epsilon(w)\, q^{-\langle w(\Lambda+\rho), \, \rho\rangle} & =  & 
q^{-\rho^2 + \sum_i i l_i}\, \prod_{i=2}^{N} \prod_{j=1}^{i-1} \bigl( 1 - q^{l_j - l_i + i - j} \bigr) \\
& = & q^{-\rho^2} q^{- \frac{1}{2} B N}\, q^{\frac{1}{2} \sum_j c_j^2}  
\prod_{i=2}^{N} \prod_{j=1}^{i-1} \bigl( 1 - q^{l_j - l_i + i - j} \bigr) \ ,
\end{eqnarray}
from which we deduce (see eq.~(\ref{1.30}))
\be\label{expp1}
\chi_\Lambda^{{\mathfrak u}(N)} (z_i) = \frac{q^{\frac{1}{2} \sum_j c_j^2}  }{V}
\prod_{i=2}^{N} \prod_{j=1}^{i-1} \bigl( 1 - q^{l_j - l_i + i - j} \bigr) \ ,
\ee
with the denominator given by the Vandermonde determinant,
\be
V  =  \sum_{w \in W} \epsilon(w)\, q^{-\langle w(\rho) - \rho, \, \rho\rangle} 
= \prod_{i=2}^{N} \prod_{j=1}^{i-1} \bigl( 1 - q^{i - j} \bigr) \ .
\ee
For a representation $\Lambda$ with finitely many boxes, the expression
appearing in  (\ref{1.31}) equals thus
\be\label{expp2}
q^{C_2} \, \dim_q \Lambda = 
q^{C_2} \, \frac{\sum_{w \in W} \epsilon(w)\, q^{-\langle w(\Lambda+\rho), \, \rho\rangle} }
{\sum_{w \in W} \epsilon(w)\, q^{-\langle w(\rho), \, \rho\rangle} }
  \cong
\frac{q^{\frac{1}{2} \sum_i r_i^2}}{V} \, \prod_{i=2}^{N} \prod_{j=1}^{i-1} \bigl( 1 - q^{l_j - l_i + i - j} \bigr) \ ,
\ee
where we have used the explicit description of the quadratic Casimir,  eq.~(\ref{A.10}), and 
dropped the $-\tfrac{B^2}{2N}$  term, because it disappears for
$N\rightarrow \infty$. At large $N$, the products in (\ref{expp1}) and (\ref{expp2}) can be written  as
\cite[Ch.~3]{Macdonald}
\be
\prod_{i=2}^{\infty} \prod_{j=1}^{i-1} \frac{ 1 - q^{r_j - r_i + i - j} }{1-q^{i-j}} 
= \prod_{(ij)\in \Lambda}(1-q^{h_{ij}})^{-1} \ ,
\ee
where the product is over boxes of the Young tableau and $h_{ij}$ is the hook length. Now we observe 
that at large $N$, (\ref{expp1}) and (\ref{expp2}) are related under transposition of the Young tableau, 
\textit{i.e.}\ upon swapping rows with columns
\be
q^{C_2} \dim_q \Lambda \cong \chi_{\Lambda^T}^{{\mathfrak u}(N)}(z_i)  \quad \quad \mbox{(boxes only)}\ .
\ee
For finite tensor powers of the antifundamental, whose Young tableaux contain only 
antiboxes, similar reasoning leads to 
\be
q^{C_2} \dim_q \Lambda \cong \chi_{\bar{\Lambda}^T}^{{\mathfrak u}(N)}(z_i)  
\quad \quad \mbox{(antiboxes only)}\ .
\ee 
In particular,  the left-hand side is therefore invariant under exchanging boxes for antiboxes.  For a 
mixed representation $\Lambda = (\bar{R}, S)$ with both boxes and antiboxes, we
finally have
\be
q^{C_2}\dim_q \Lambda \cong \chi_{R^T}^{{\mathfrak u}(N)}(z_i)\, 
 \chi_{S^T}^{{\mathfrak u}(N)}(z_i)  \ .
\ee

\section{Fusion Calculation Details}\label{a:fusiondetails}

In this section we provide some more details of the fusion calculations. 

\subsection{The Calculation for $({\rm f};0)\otimes (0;{\rm f})$}

Let us explain in more detail the analysis leading to the description of 
${\cal H}^{(1)}\equiv ({\cal H}_1 \otimes {\cal H}_2)^{(1)}$ for the case of the fusion 
$({\rm f};0)\otimes (0;{\rm f})$. For the analysis at level one, we are only allowed to divide
by states that are in the image of modes whose total mode number is smaller than $-1$. So 
in particular, we have the relations 
\be
\begin{array}{lrl}
0\cong W_{-2} (\phi_1\otimes \phi_2): \quad & 
(L_{-1}^2\phi_1) \otimes \phi_2 & \cong \phi_1 \otimes ( L_{-1}^2\phi_2) \\
0\cong U_{-3} (\phi_1\otimes \phi_2): \quad & 
(L_{-1}^3\phi_1) \otimes \phi_2 & \cong - \phi_1 \otimes ( L_{-1}^3\phi_2) \\
0\cong U_{-2} (\phi_1\otimes \phi_2): \quad & 
(L_{-1}^3\phi_1) \otimes \phi_2 & \cong - \tfrac{1}{2} (3+\lambda) (L_{-1}^2\phi_1) \otimes \phi_2
- \tfrac{1}{2}  (3-\lambda)  \phi_1 \otimes ( L_{-1}^2\phi_2) \\
0\cong L_{-1} L_{-1} (\phi_1\otimes \phi_2): 
& (L_{-1}^2\phi_1) \otimes \phi_2 & \cong  - 2 (L_{-1}\phi_1) \otimes (L_{-1}\phi_2) - 
\phi_1 \otimes ( L_{-1}^2\phi_2) \ ,
\end{array}
\ee
where we have used the various null vector identities.
In particular, combining the first and third relation, as well as the first and last relation, we then obtain
\begin{eqnarray}
\label{A1}
(L_{-1}^3\phi_1) \otimes \phi_2  & \cong & - 3 (L_{-1}^2\phi_1) \otimes \phi_2 \\
(L_{-1} \phi_1) \otimes (L_{-1} \phi_2) & \cong & - (L_{-1}^2\phi_1) \otimes \phi_2  
\cong - \phi_1 \otimes (L_{-1}^2\phi_2) \ .\notag
\end{eqnarray}
The space at level one is therefore spanned by (\ref{4.26}), and we find for the action of $L_0$,
\begin{eqnarray}
L_0 f_1 & = & f_1 + f_2 \\
L_0 f_2 & = & 2 f_2 + f_4 \notag\\
L_0 f_3 & = & 2 f_3 - f_4 \notag\\
L_0 f_4 & = & 3 f_4 - 3 f_4 = 0 \ ,\notag
\end{eqnarray}
leading to the eigenvectors given in (\ref{4.28}). 
Using (\ref{4.30}) we can finally determine the action of $L_1$ on these states,
\begin{eqnarray}
L_1 f_1 & = & (1+\lambda) f_1 + f_2 \\
L_1 f_2 & = &  (3+\lambda) f_2 +(1+\lambda) f_1+ f_4 \cong (1+\lambda) (e_1 + e_2) \notag\\
L_1 f_3 & = &   (1+\lambda) f_3 + (1-\lambda) f_1 - f_4 \cong (1-\lambda) (e_1 + e_2) \notag\\
L_1 f_4 & = &  (5+\lambda) f_4 + 2 (\lambda+2) f_2 - 3 e_4 \cong 0 \ ,\notag
\end{eqnarray}
where we have implemented the relations $f_3 = - e_2$ and $f_4 = - 2 e_2$ since the image is to be 
interpreted in ${\cal H}^{(0)}\equiv ({\cal H}_1 \otimes {\cal H}_2)^{(0)}$. 
On the $L_0$ eigenstates this then leads to (\ref{4.31}).

\subsection{Another Example}\label{sec:4.2.1}
Now consider $({\rm f};0)\otimes (0;\tableau{1 1})$. In this case, the space at level one is spanned by the five vectors
\be\label{H1s}
f_1 = \phi_1 \otimes \phi_2 \ , \ f_2 = (L_{-1}\phi_1)\otimes \phi_2  \ , \ 
f_3 = \phi_1 \otimes (L_{-1}\phi_2) \ , \ 
f_4 = (L_{-1}^2 \phi_1) \otimes \phi_2 \ , \ 
f_5 = \phi_1 \otimes (L_{-1}^2 \phi_2)  \ ,
\ee
and we have the relations 
\begin{eqnarray}
(L_{-1}^3 \phi_1) \otimes \phi_2 & \cong & - 3 f_4 \\
(L_{-1}^2 \phi_1) \otimes (L_{-1}\phi_2) & \cong & 2 f_4 + f_5\notag\\ 
(L_{-1} \phi_1) \otimes (L_{-1}^2\phi_2) & \cong & - f_4 - 2 f_5\notag\\ 
(L_{-1}\phi_1)\otimes (L_{-1}\phi_2) & \cong  & - \tfrac{1}{2} f_4  - \tfrac{1}{2} f_5 \ ,\notag
\end{eqnarray}
where we have used the null relations for $\phi_2$ given in (\ref{asymnul}).
Furthermore, in order to go to the highest weight space
${\cal H}^{(0)} \equiv ({\cal H}_1 \otimes {\cal H}_2 )^{(0)}$ 
we have to impose the additional relations
\be\label{nur}
{\cal H}^{(0)} : \qquad f_3 = - f_2 \ , \qquad  f_5 = f_4 = - 2 f_2 \ .
\ee
We can determine the action of $L_0$ on this five-dimensional space, and we find the eigenvectors
\be
\begin{array}{ll}
h=\tfrac{1}{2} (1-\lambda): \qquad \qquad & \mu^{(1)} = - \tfrac{1}{2}\, f_4 \\
h=\tfrac{1}{2} (3-\lambda) : \qquad & \nu^{(1)} = f_1 - f_2 - f_4\\
& \rho^{(1)} = \tfrac{1}{2} f_4 -\tfrac{1}{2} f_5 \\
h=\tfrac{1}{2} (5-\lambda) : \qquad &
\sigma^{(1)} = 2 f_3 - f_5 \\
& \pi^{(1)} = 2 f_2 + f_4 \ . 
\end{array}
\ee
It follows from (\ref{nur}) that $\mu$ and $\nu$ are the states that survive in the `highest weight space', whereas
$\rho$, $\sigma$ and $\pi$ are descendants. 
Because there are now multiplicities, we have used the action of $W_0$ to find those linear combinations
that are also $W_0$ eigenstates; in fact we find
\be\label{W0ei}
\begin{array}{rcl}
W_0 \, \nu^{(1)} & = &  (\lambda^2 - 9 \lambda + 2)\,  \nu^{(1)} \\
W_0\, \rho^{(1)} & = & (2-\lambda) (7-\lambda)\, \rho^{(1)} \ , 
\end{array}
\ee
and thus the eigenstates are uniquely fixed by this requirement. 
Finally, we apply again $L_1$ to these states and find 
\be
\begin{array}{rclrcl}
L_1 \mu^{(1)} & = & 0 \qquad \qquad & L_1 \rho^{(1)} & = &  (1-\lambda) \mu^{(0)}   \\
L_1 \nu^{(1)} & = & - \lambda \mu^{(0)} & L_1 \sigma^{(1)} & = & 4 (1-\lambda) \nu^{(0)} 
\end{array}
\qquad \qquad
L_1 \pi^{(1)}  =  2 (1+\lambda) \nu^{(0)} \ . 
\ee
Thus the representation is indeed again indecomposable for $\lambda\neq 0$ since $L_1\nu\sim \mu$.

\subsection{The Double Blind Test}\label{ss:doubleblind}

We have also checked that in situations where we expect the answer to be a direct sum, this analysis
also predicts it to be a direct sum (for any value of $\lambda$). 
In particular, we have done the analysis for 
$({\rm f};0)\otimes ({\rm f};0)$ for which the space at level one is four-dimensional, and spanned by 
\be\label{H1ff}
f_1 = \phi_1 \otimes \phi_2 \ , \ f_2 = (L_{-1}\phi_1)\otimes \phi_2  \ , \ 
f_3 = \phi_1 \otimes (L_{-1}\phi_2) \ , \ 
f_4 = (L_{-1}^2 \phi_1) \otimes \phi_2 \ .
\ee
The $L_0$ eigenvectors turn out to be in that case
\be
\begin{array}{ll}
h=1+\lambda: \qquad \qquad & \omega^{(1)} = f_1-f_2+\tfrac{1}{2}f_4 \\
h=2+\lambda: \qquad & \psi^{(1)} = f_2 - f_3 - f_4 \\
& \rho^{(1)} = f_2+ f_3-f_4 \\
h=3+\lambda: \qquad &  \xi^{(1)} = f_4 \ ,
\end{array}
\ee
but now the eigenvectors are not uniquely fixed because both $\psi^{(1)}$ and
$\rho^{(1)}$ have the same $W_0$ eigenvalue, namely $-2 (2+\lambda)(4+\lambda)$. There is therefore
one linear combination of the two states at conformal weight $h=2+\lambda$ that is in fact annihilated
by $L_1$; in the above conventions it is the state $\psi^{(1)}$, {\it i.e.}\ we have $L_1\psi^{(1)}=0$. Thus
unlike the other cases above, the representation appears to be a direct sum of two representations,
in agreement with our expectations. 

\subsection{A Higher Order Example}\label{sec:4.2.3}

Finally, we have studied the example $(\tableau{1 1} ;0) \otimes (0; \tableau{1 1})$.
Writing $\phi_1 = (\tableau{1 1} ;0)$ and  $\phi_2 = (0; \tableau{1 1})$, 
the eigenvalues are 
\be
\begin{array}{ll}
h_1 = 1+\lambda \qquad & h_2 = 1 -\lambda \\
w_1 = - 2(1+\lambda)(2+\lambda) \qquad & w_2 = 2(1-\lambda)(2-\lambda) \\
u_1 = 2(1+\lambda)(2+\lambda)(3+\lambda) \qquad & u_2 = 2(1-\lambda)(2-\lambda)(3-\lambda) \\
x_1 = - 2 (1+\lambda)(2+\lambda)(3+\lambda)(4+\lambda) \qquad\qquad 
& x_2 = 2 (1-\lambda)(2-\lambda)(3-\lambda)(4-\lambda) \ .
\end{array}
\ee
Here $x_i$ is the eigenvalue of the spin $5$ field $X$ with commutation relations
\be
{}[L_m,X_n ] = (4m-n) X_{m+n} \qquad  [X_m,L_n] = (m - 4n) X_{m+n}  \ . 
\ee
We have the null-relations
\be
\begin{array}{ll}\label{asymnul}
W_{-1}\phi_1 = - 3 (2+\lambda) L_{-1} \phi_1 \qquad 
& W_{-1}\phi_2 =  3 (2-\lambda) L_{-1} \phi_2 \\
U_{-1} \phi_1 = 4 (2+\lambda)(3+\lambda) L_{-1} \phi_1 \qquad
& U_{-1} \phi_2 = 4 (2-\lambda)(3-\lambda) L_{-1} \phi_2 \\
X_{-1} \phi_1 = -5 (2+\lambda)(3+\lambda)(4+\lambda) L_{-1} \phi_1 \qquad \qquad & 
X_{-1} \phi_2 =  5 (2-\lambda)(3-\lambda)(4-\lambda) L_{-1} \phi_2 \\
U_{-2} \phi_1 = - \tfrac{5}{3} (3+\lambda) W_{-2} \phi_1 \qquad \qquad & 
U_{-2} \phi_2 =  \tfrac{5}{3} (3-\lambda) W_{-2} \phi_2 \\
X_{-2} \phi_1 =  \tfrac{5}{2} (3+\lambda) (4+\lambda) W_{-2} \phi_1 \qquad \qquad & 
X_{-2} \phi_2 =  \tfrac{5}{2} (3-\lambda) (4-\lambda) W_{-2} \phi_2 \\
X_{-3} \phi_1 = - \tfrac{7}{4} (4+\lambda) U_{-3} \phi_1 \qquad \qquad & 
X_{-3} \phi_2 =  \tfrac{7}{4}  (4-\lambda) U_{-3} \phi_2 \ ,
\end{array}
\ee
as well as the two relations\footnote{Some of these null relations are fixed by the action of $L_1$, while others were derived assuming the conjecture of \cite{Gaberdiel:2011wb} relating $\W_N$ to the wedge algebra of \w{\lambda}.}
\be\label{U3}
\Bigl(U_{-3} + 5 L_{-1} W_{-2} + 10 L_{-1}^3 \Bigr) \phi_1 = 
\Bigl(U_{-3} - 5 L_{-1} W_{-2} + 10 L_{-1}^3 \Bigr) \phi_2 =  0 
\ee
and
\be
\Bigl( L_{-1}^4 - \tfrac{1}{12} W_{-2} W_{-2} - \tfrac{1}{5} L_{-1} U_{-3} - \tfrac{3}{35} X_{-4} \Bigr) \phi_1  = 0  
\ee
and
\be
\Bigl( L_{-1}^4 - \tfrac{1}{12} W_{-2} W_{-2} - \tfrac{1}{5} L_{-1} U_{-3} + \tfrac{3}{35} X_{-4} \Bigr) \phi_2  = 0  \ .
\ee

\subsubsection{The highest weight space}

For the highest weight space we have the relations
\be
\begin{array}{llcl}
L_{-1}: \qquad & (L_{-1}\otimes {\bf 1}) &\cong & - ({\bf 1}\otimes L_{-1}) \\
W_{-2}: \qquad & (W_{-2}\otimes {\bf 1}) &\cong & - ({\bf 1}\otimes W_{-2}) \\
W_{-1}: \qquad & (W_{-2}\otimes {\bf 1}) &\cong & 12 (L_{-1} \otimes {\bf 1}) \\
U_{-3}: \qquad & (U_{-3}\otimes {\bf 1}) &\cong & - ({\bf 1}\otimes U_{-3}) \\
U_{-2}: \qquad & (U_{-3}\otimes {\bf 1}) &\cong & 120  (L_{-1} \otimes {\bf 1}) \ ,
\end{array}
\ee
where we have used the short-hand notation $(S_1\otimes S_2) \equiv (S_1 \phi_1 \otimes S_2 \phi_2)$,
and indicated on the left where the relation originally comes from. From 
$W_{-1}$ applied to the state  $({\bf 1}\otimes L_{-1})$ we obtain, using the above relations
as well as the null vector relations for $W_{-1}$
\be
(W_{-2}\otimes L_{-1}) \cong 12 (L_{-1}\otimes {\bf 1}) - 12 (L_{-1}^2 \otimes {\bf 1}) \ ,
\ee
and the null vector (\ref{U3}) gives rise to the relation 
\be
120  (L_{-1} \otimes {\bf 1}) \cong 5 (W_{-2} \otimes L_{-1}) - 10 (L_{-1}^3 \otimes {\bf 1}) \ .
\ee
Combing the last two equations we conclude that
\be
(L_{-1}^3\otimes {\bf 1}) \cong - 6 (L_{-1}^2 \otimes {\bf 1}) - 6 (L_{-1} \otimes {\bf 1}) \ .
\ee
The highest weight space is therefore spanned by 
\be\label{4.46}
{\cal H}^{(0)} = \hbox{span} \bigl\{ 
e_1 = \phi_1 \otimes \phi_2 \ , \qquad e_2 = (L_{-1} \phi_1) \otimes \phi_2 \ , \qquad
e_3 = (L_{-1}^2 \phi_1) \otimes \phi_2 \bigr\} \ .
\ee
For the action of $L_0$ on these states we find 
\begin{eqnarray}
L_0 e_1 & = & 2 e_1 + e_2 \\
L_0 e_2 & = & 3 e_2 + e_3 \notag\\
L_0 e_3 & = & -2 e_3 - 6 e_2 \ ,\notag
\end{eqnarray}
leading to the eigenvectors 
\be\label{4.47}
\begin{array}{llcl}
h=0: \qquad \qquad & \omega^{(0)} & = & e_3 + 2 e_2 \\
h=1: \qquad & \psi^{(0)} & = & e_3 + 3 e_2 \\
h=2: \qquad & \rho^{(0)} & = & 2 e_1 + 4 e_2 + e_3 \ . 
\end{array}
\ee
The $W_0$ action is 
\begin{eqnarray}
W_0 e_1 & = & -12 \lambda e_1 -6\lambda  e_2 \\
L_0 e_2 & = & -18\lambda e_2 -6\lambda e_3 \notag\\
L_0 e_3 & = & 12 \lambda e_3 + 36\lambda e_2 \ ,\notag
\end{eqnarray}
from which one concludes that
\be\label{W00}
W_0 \omega^{(0)} = 0 \ , \qquad
W_0 \psi^{(0)} = - 6 \lambda \psi^{(0)} \ , \qquad
W_0 \rho^{(0)} = - 12 \lambda \rho^{(0)} \ .
\ee

\subsubsection{The analysis at the first excited level}

In ${\cal H}^{(1)}$ we have the relations
\be
\begin{array}{lrcl}
L_{-1}^2 : \qquad
& (L_{-1}\otimes L_{-1}) & \cong & - \tfrac{1}{2} (f_4+f_6) \\
W_{-2} : \qquad
& ({\bf 1}\otimes W_{-2}) & \cong & - f_5 \\
L_{-1}W_{-1} : \qquad \qquad
& (L_{-1}W_{-2}\otimes {\bf 1}) & \cong &  - (W_{-2}\otimes L_{-1}) + 6 f_4 - 6 f_6 \ .
\end{array}
\ee
Using null vectors we deduce in addition
\be
\begin{array}{rcl}
(U_{-3}\otimes {\bf 1}) & \cong & 10 (W_{-2}\otimes {\bf 1}) \\
(L_{-1}W_{-2}\otimes {\bf 1}) & \cong & - 2 f_5 - 2 f_7 \\
(W_{-2}\otimes L_{-1}) & \cong & 6 f_4 + 2 f_5 - 6 f_6 + 2 f_7 \\
(L_{-1}^2 \otimes L_{-1}) & \cong & - f_4 + f_6 - f_7 \\
(L_{-1} \otimes L_{-1}^2) & \cong &  2 f_4 - 2 f_6 + f_7 \\
(X_{-4} \otimes {\bf 1}) & \cong & 140 f_5 \\
(W_{-2} W_{-2} \otimes {\bf 1}) & \cong & - 144 f_4 - 72 f_5 - 48 f_7 \\
(U_{-3} L_{-1} \otimes {\bf 1}) & \cong & - 30 f_5 - 20 f_7 \\
(L_{-1}^4\otimes {\bf 1}) & \cong & - 12 f_4 - 8 f_7 \ ,
\end{array}
\ee
where we have defined the $f_i$ by 
\be\label{4.49}
\begin{array}{lcllcl}
f_1 & = & (\phi_1\otimes \phi_2) \qquad \qquad 
& f_2 & = & (L_{-1}\phi_1\otimes \phi_2)  \\
f_3 & = & (\phi_1\otimes L_{-1} \phi_2) \qquad \qquad 
& f_4 & = & (L_{-1}^2\phi_1\otimes \phi_2)  \\
f_5 & = & (W_{-2}\phi_1\otimes \phi_2) \qquad \qquad 
& f_6 & = & (\phi_1\otimes L_{-1}^2\phi_2)  \\
f_7 & = & (L_{-1}^3\phi_1\otimes \phi_2)  \ .& &
\end{array}
\ee
These states form then a basis for ${\cal H}^{(1)}$. Note that 
 in ${\cal H}^{(0)}$ we have the relations
\be
{\cal H}^{(0)}: \quad f_1 = e_1 \ , \quad f_2 = - f_3 = e_2 \ , \quad f_5 = 12 e_2 \ , \qquad
f_4 = f_6 = e_3 \ , \qquad f_7 = - 6 e_2 - 6 e_3 \ .
\ee 
The action of $L_0$ on these basis elements is 
\begin{eqnarray}
L_0 f_1 & = & 2 f_1 + f_2 \\
L_0 f_2 & = & 3 f_2 + f_4 \notag \\
L_0 f_3 & = & 3 f_3 - \tfrac{1}{2} f_4 - \tfrac{1}{2} f_6 \notag\\
L_0 f_4 & = & 4 f_4 + f_7 \notag\\
L_0 f_5 & = & 2 f_5 - 2 f_7 \notag\\
L_0 f_6 & = & 2 f_4 + 2 f_6 + f_7 \notag\\
L_0 f_7 & = & - 12 f_4 - 3 f_7 \ ,\notag
\end{eqnarray}
from which we determine the eigenvectors to be 
\be\label{4.50}
\begin{array}{llcl}
h=0: \qquad \qquad & \omega^{(1)} & = & 3 f_4 + f_7 \\
h=1: \qquad & \psi^{(1)} & = & 4 f_4 + f_7 \\
h=2: \qquad & \rho^{(1)} & = & 12 (f_1-f_2) - 18 f_4 + f_5 - 4 f_7  \\
		    & \sigma^{(1)} & = & 12 f_4 + f_5 + 2 f_7 \\
		    & \xi^{(1)} & = & 6f_4 + f_5 + 6f_6 + 2f_7 \\
h=3: \qquad & \pi^{(1)} & = & -2 f_2 - 2 f_3 - f_4 + f_6 \\
		    & \chi^{(1)} & = & 6 f_2 + 6 f_4 + f_7 \ . \
\end{array}
\ee
In order to determine the action of $W_0$ on these vectors, we use that (from the
highest weight analysis) the eigenvalues of $\omega$ and $\psi$ under $W_0$ are known.
This then implies that 
\begin{eqnarray}
W_0 f_4 & = & -24 \lambda f_4 - 6 \lambda f_7 \\
W_0 f_7 & = & 72\lambda f_4 + 18\lambda f_7 \ . \notag
\end{eqnarray}
We have also worked out from first principles
\begin{eqnarray}
W_0 f_1 & = & -12 \lambda f_1 - 6 (2+\lambda) f_2 + f_5 \\
W_0 f_2 & = & - 6 (2+3\lambda) f_2 - 6 (2+\lambda) f_4 - 2 f_7 \notag\\
W_0 f_3 & = & 6(2-3\lambda) f_3 + 3(4+\lambda)f_4 + 2 f_5 + 3\lambda f_6 + 2 f_7\notag\\
W_0 f_5 & = & -144 f_4 - 12 (1+\lambda) f_5 - 12 (2-\lambda) f_7 \notag\\
W_0 f_6 & = & 12(3-\lambda)f_4 + 4 f_5 + 12(1-\lambda)f_6 + 2(4-3\lambda)f_7 \ ,\notag
\end{eqnarray}
which allows us to check that 
\be\label{W001}
W_0 \rho = - 12 \lambda \rho \ , \qquad
W_0 \sigma =  - 12 (\lambda+1) \sigma \ , \quad W_0 \xi = 12(1-\lambda)\xi \ .
\ee
Thus $\rho$ and $\sigma$ are the (unique) eigenvectors of $W_0$ at $h=2$. 
Finally, we have worked out the action of $L_1$ on these states, finding
\begin{eqnarray}
L_1 f_1 & = & (2+2\lambda) e_1 + e_2 \\
L_1 f_2 & = & (2+2\lambda) e_1 + (4+2\lambda) e_2 + e_3 \notag\\
L_1 f_3 & = & (2-2\lambda) e_1 - (2+2\lambda) e_2 - e_3 \notag\\
L_1 f_4 & = & 4 \lambda e_2 + 2 \lambda e_3 \notag\\
L_1 f_5 & = & 12 (3+\lambda) e_2 + 12 e_3 \notag\\
L_1 f_6 & = & 4 (\lambda-3) e_2 + 2 (\lambda-2) e_3 \notag\\
L_1 f_7 & = & -12 \lambda e_2 - 6 \lambda e_3 \ .\notag
\end{eqnarray}
This leads to 
\be\label{4.53}
\begin{array}{rclrcl}
L_1 \omega^{(1)} & = & 0 \qquad \qquad \qquad & L_1 \psi^{(1)} & = & 2 \lambda \omega^{(0)}  \\
L_1 \sigma^{(1)} & = &  12 (1+\lambda) \psi^{(0)} & 
L_1 \rho^{(1)} & = & -12 \lambda \psi^{(0)} 
\end{array}
\qquad \qquad 
L_1 \xi^{(1)}  =  12(\lambda-1) \psi^{(0)} \ . 
\ee
Thus, provided that $\lambda\neq 0$, $L_1$ maps $\rho^{(1)}$ to $\psi^{(0)}$ and 
$\psi^{(1)}$ to $\omega^{(0)}$, thus demonstrating that the representation is indecomposable, 
with cyclic vector $\rho$. The resulting structure is sketched in (\ref{ffdiac}).

\section{Zero Modes of Degenerate Representations}\label{a:hsdetails}

In this section we check the formula (\ref{wsid}) for spins $s=3,4$.  This formula equates 
the zero mode eigenvalues of degenerate representations of the \hs{\lambda} algebra to those 
of $\W_N$ in the 't~Hooft limit.  The higher spin zero modes are a straightforward extension of 
the conformal dimensions.  The one new ingredient is that higher spin generators are ambiguous, 
so to make the comparison we must first go to a basis where the generators are primary.  A closed 
form expression for the zero modes is known only in a particular non-primary basis (the Miura basis) 
with currents $U^s$ related to $W^s$ by a nonlinear field redefinition.  The non-primary zero modes 
eigenvalues are \cite{Bilal:1991eu,Bouwknegt:1992wg}
\be\label{uvalue}
u_s(\Lambda) =  (-1)^{s-1}\sum_{i_1 <\dots < i_s} \prod_{j=1}^s [(\Lambda, h_{i_j}) + (s-j)\alpha_0]  \ .
\ee
The transformation to the primary basis through spin 4 is \cite{DiFrancesco:1990qr}
\bea
W^2(z) &=& U^2(z) \\
W^3(z) &=& U^3(z) - \frac{N-2}{2}\alpha_0\p U^2(z) \notag\\
W^4(z) &=& U^4(z) - \frac{N-3}{2}\alpha_0 \p U^3(z) + \frac{(N-2)(N-3)}{10}\alpha_0^2 \p^2 U^2(z) \notag\\
& & \quad - \frac{(N-2)(N-3)(5N+7)}{10N(N^2-1)}(U^2(z))^2 \ . \notag
\eea
Converting to modes, performing the sums in (\ref{uvalue}) at finite $N$ and $\alpha_0$, and taking the 
two separate limits (\ref{tHooftalpha}) and (\ref{hsalpha}), respectively, we find
\begin{align}
\mbox{'t Hooft:} \quad & w^{(3)}_{\rm mm}(\Lambda) = \frac{\lambda^2}{6}B   + \frac{\lambda}{2} \sum r_i^2 + \frac{1}{3}\sum r_i^3\\
& w^{(4)}_{\rm mm}(\Lambda) = \frac{\lambda(1+\lambda^2)}{20}B + \frac{1+6\lambda^2}{20}\sum r_i^2 + \frac{\lambda}{2}\sum r_i^3 + \frac{1}{4}r_i^4 \notag\\
\mbox{Higher spin:} \quad & w^{(3)}_{\rm hs}(\Lambda)k_{\rm DS}^{-1/2} = -\sum i^2r_i + (1-\lambda)\sum ir_i - \frac{B}{6}(1-\lambda)(2-\lambda)\notag\\
& w^{(4)}_{\rm hs}(\Lambda)k_{\rm DS}^{-1} = \sum i^3r_i + \frac{3(\lambda-1)}{2}\sum i^2 r_i + \frac{11-15\lambda+6\lambda^2}{10}\sum ir_i \notag\\
& \hspace{4cm}- \frac{B}{20}(1-\lambda)(2-\lambda)(3-\lambda) \ .\notag
\end{align}
Using the Young tableau identities
\be
\sum i r_i = \frac{1}{2} \sum c_i(c_i+1)  \ , \quad
\sum i^2 r_i = \frac{1}{6} \sum c_i(1+c_i)(1+2c_i)\ , \quad
\sum i^3 r_i = \frac{1}{4}\sum c_i^2(1+c_i)^2 
\ee
we find (up to an unfixed $\lambda$-independent normalization)
\be
w^{(s)}_{\rm mm}(\Lambda) = w^{(s)}_{\rm hs}(\Lambda^T) \ .
\ee
Therefore the degenerate representations of the two algebras match after transposing the Young tableaux, 
at least through spin 4. 

\begin{singlespace}

\bibliographystyle{JHEP}
\bibliography{references}
\end{singlespace}

\end{document}